\documentclass[conference,compsoc]{IEEEtran}%
\IEEEoverridecommandlockouts
\usepackage{amsmath,amsfonts}
\usepackage{algorithmic}
\usepackage{array}
\usepackage[caption=false,font=normalsize,labelfont=sf,textfont=sf]{subfig}
\usepackage{textcomp}
\usepackage{stfloats}
\usepackage{url}
\usepackage{verbatim}
\usepackage{hhline}
\usepackage{graphicx}
\usepackage{adjustbox}
\usepackage{makecell}
\usepackage{multirow}

\usepackage{hyperref}
\hypersetup{
	colorlinks=true,
	linkcolor=red,
	filecolor=magenta,      
	urlcolor=cyan,
	pdftitle={Overleaf Example},
	pdfpagemode=FullScreen,
}

\usepackage{arydshln}

\usepackage[table,xcdraw]{xcolor}

\def\BibTeX{{\rm B\kern-.05em{\sc i\kern-.025em b}\kern-.08em
		T\kern-.1667em\lower.7ex\hbox{E}\kern-.125emX}}
\usepackage{balance}

\begin{document}
	\title{Prompt Engineering-assisted Malware Dynamic Analysis Using GPT-4}

\author{
	\IEEEauthorblockN{\textbf{\textit{Pei Yan}}}
	\IEEEauthorblockA{\textsuperscript{1}\textit{Guangdong Key Laboratory of } \\
		\textit{Intelligent Information Processing} \\
		\textsuperscript{2}\textit{Shenzhen Key Laboratory of Media Security}\\
		Shenzhen, China \\
		{yanpei2022@email.szu.edu.cn}}\\

	\IEEEauthorblockN{\textbf{\textit{Miaohui Wang}}}
	\IEEEauthorblockA{\textsuperscript{1}\textit{Guangdong Key Laboratory of } \\
	\textit{Intelligent Information Processing} \\
	\textsuperscript{2}\textit{Shenzhen Key Laboratory of Media Security}\\
	Shenzhen, China \\
	wang.miaohui@gmail.com}\\
	
	\and
	
	\IEEEauthorblockN{\textbf{\textit{Shunquan Tan}}}
	\IEEEauthorblockA{\textsuperscript{1}\textit{Guangdong Key Laboratory of } \\
	\textit{Intelligent Information Processing} \\
	\textsuperscript{2}\textit{Shenzhen Key Laboratory of Media Security}\\
	Shenzhen, China \\
	tansq@szu.edu.cn}\\

	\IEEEauthorblockN{\textbf{\textit{Jiwu Huang$^{\ast}$ }}}
	
	\IEEEauthorblockA{\textsuperscript{1}\textit{Guangdong Key Laboratory of } \\
	\textit{Intelligent Information Processing} \\
	\textsuperscript{2}\textit{Shenzhen Key Laboratory of Media Security}\\
	Shenzhen, China \\
	jwhuang@szu.edu.cn}
\thanks{*Corresponding author}

}
	
	\maketitle
	
	\begin{abstract}
Dynamic analysis methods effectively identify shelled, wrapped, or obfuscated malware, thereby preventing them from invading computers. As a significant representation of dynamic malware behavior, the API (Application Programming Interface) sequence, comprised of consecutive API calls, has progressively become the dominant feature of dynamic analysis methods. 
Though there have been numerous deep learning models for malware detection based on API sequences, the quality of API call representations produced by those models is limited. These models cannot generate representations for unknown API calls, which weakens both the detection performance and the generalization. Further, the concept drift phenomenon of API calls is prominent. 
To tackle these issues, we introduce a prompt engineering-assisted malware dynamic analysis using \textit{GPT-4}. In this method, \textit{GPT-4} is employed to create explanatory text for each API call within the API sequence. Afterward, the pre-trained language model BERT (Bidirectional Encoder Representations from Transformers) is used to obtain the representation of the text, from which we derive the representation of the API sequence. Theoretically, this proposed method is capable of generating representations for all API calls, excluding the necessity for dataset training during the generation process. Utilizing the representation, a CNN-based detection model is designed to extract the feature. 
We adopt five benchmark datasets to validate the performance of the proposed model. The experimental results reveal that the proposed detection algorithm performs better than the state-of-the-art method (TextCNN). Specifically, in cross-database experiments and few-shot learning experiments, the proposed model achieves excellent detection performance and almost a 100\% recall rate for malware, verifying its superior generalization performance.
The code is available at: \href{https://github.com/yan-scnu/Prompted_Dynamic_Detection}{\tt{github.com/yan-scnu/Prompted\_Dynamic\_Detection}}.
	\end{abstract}
	
	\begin{IEEEkeywords}
Computer security, malware detection, prompt engineering, large language model
	\end{IEEEkeywords}

	\section{Introduction}
	\begin{figure}[t]
		
		\begin{minipage}{0.5\textwidth}
			\centering
			\includegraphics[width=1\linewidth]{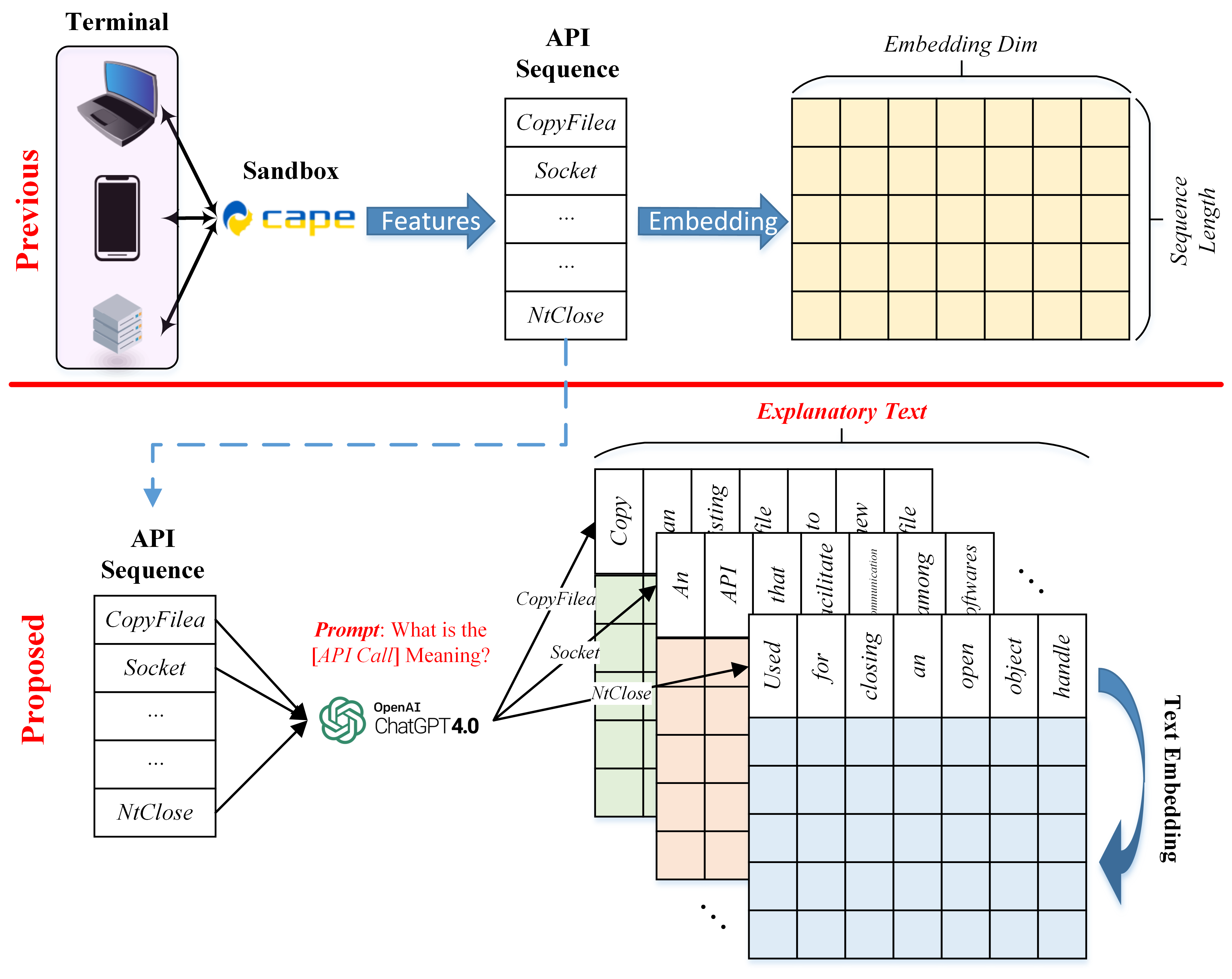} 
			\caption{The comparison between the proposed and the existing representation methods. When placing the suspicious code from the terminal into a sandbox for execution,  the sandbox will output the API sequence called by code. We propose to employ \textit{GPT-4} for generating explanatory text for each API call within the API sequence. We then embed this explanatory text and concatenate the resulting embedded matrices to achieve representation. By introducing additional prior knowledge, our method enhances representation efficacy, thereby outperforming previous methods.}
			\label{Intro}
		\end{minipage}
		
	\end{figure}
Malware poses a significant risk to the network and computer security, rendering effective malware detection an essential and challenging task\cite{guizani2020network,amira2023survey,gopinath2023comprehensive}. A dynamic analysis method was proposed to achieve better detection performance. This method involves running the malware in a sandbox and recording its dynamic behavior. The code is judged based on its dynamic behavior, providing a better mechanism to detect obfuscated and packaged malware. 

For dynamic analysis methods, the most commonly used dynamic behavior is the API sequence, composed of API calls during the code's runtime process. The encoded API sequence shares certain similarities with encoded natural text\cite{uppal2014malware,pascanu2015malware}. As Deep Learning (DL) technology has yielded promising results in Natural Language Processing (NLP) tasks, many DL-based text classification models have been applied to identifying API sequences, achieving excellent classification performance\cite{athiwaratkun2017malware,maniath2017deep,catak2020deep}. With the recent rise of the Large Language Model (LLM)\cite{vaswani2017attention,devlin2018bert,radford2019language}, Many fields are attempting to integrate large language models to solve problems \cite{radford2021learning,yao2024PromptCARE}. Malware analysts are exploring ways to further enhance detection performance using LLMs. For instance, Transformer-based network architectures are being designed for malware classification\cite{demirkiran2022ensemble,xu2021malbert}.

However, due to limited training data and the differences between API sequences and encoded text sequences, the detection performance of state-of-the-art(SOTA) methods remains constrained. Moreover, several issues persist in current API-based detection models.
\begin{itemize}
\item \textbf{Limited representation of malware features.} The representation of API calls is refined during the training process. The quality of the training data significantly influences the effectiveness of the representation.
\item \textbf{Weak generalization performance.} The representation, obtained during the training phase, may be susceptible to overfitting on a specific dataset. Consequently, this can lead to inferior performance when validated against other datasets.
\item \textbf{Sensitive to concept drift.} As systems and detection tools iterate, API calls will also be updated accordingly. This can cause the current detection model to lack representation for new API calls. Consequently, it may affect the detection accuracy of future malware.
\end{itemize}

To address the issues mentioned above, this paper introduces a method for generating representation based on a LLM. This approach employs LLM to produce explanatory text for each API call. We then perform embedding operations on these texts, which serve as the representations for the API calls. With the high-quality explanatory text created by the LLM and the application of pre-trained models, we can directly obtain the representation. This eliminates the need for training with the API sequence dataset, significantly enhancing the efficiency of representation generation.
 
\textit{GPT-4} \cite{GPT4} had demonstrated its better performance than other models (\textit{e.g.}, PaLm \cite{chowdhery2022palm}, Llama \cite{touvron2023llama},
ChatGLM \cite{chatglm}) in various language tasks. Consequently, this study employs \textit{GPT-4} for the generation of explanatory text for API calls. We generate representations based on this explanatory text and concatenate the text's representations to obtain the API sequence's representation, as illustrated in Figure~\ref{Intro}. We then design appropriate deep neural networks to learn these representations and classify malware based on them. The primary contributions of this paper are summarized as follows: 
	
	\begin{enumerate}
		\item We guide \textit{GPT-4} to generate explanatory texts for API calls, and these texts serve as a representation of each API call during both training and testing procedures. To the best of our knowledge, this is the first report to apply prompt engineering to dynamic malware analysis.
		
		\item With the assistance of descriptive explanatory text, the acquisition of API call representations does not require training with datasets. This approach introduces the representation with more additional knowledge, thereby improving the quality of the representations and enhancing their generalizability.
		
		\item Thanks to the substantial training corpus and robust text restatement capabilities of the \textit{GPT-4}, it can theoretically generate the representation for all API calls. This not only makes the representation association denser, stronger, and more stable, but also benefits the coping mechanisms for data drift phenomena such as API call updates.

	\end{enumerate}

	\section{Related Work}
	\subsection{Malware Dynamic Analysis}
	
	The dynamic analysis method analyzes malware by examining its dynamic behavioral features during execution \cite{alazab2010towards}. By executing the executable file in a sandbox (such as Cuckoo\footnote{https://github.com/cuckoosandbox/cuckoo} or Cape\footnote{https://github.com/mandiant/capa}), the sandbox records its behavioral characteristics, which enables us to analyze the file based on these attributes. In comparison with static analysis methods, which do not require execution during the analysis process, dynamic analysis methods can effectively detect shell, obfuscation, and packaged malware. As such, it is currently one of the primary detection methods. 

One of the most crucial dynamic behavioral characteristics is the API sequence \cite{uppal2014malware,pascanu2015malware}. The API sequence reflects the interaction between the code and the operating system, providing an understanding of the malware behaviors, which is essential for designing effective malware defense strategies. Earlier research leveraged statistical learning \cite{gupta2016malware,ravi2012malware,ki2015novel}, machine learning \cite{sami2010malware,pektacs2018malware}, and graph methods \cite{anderson2011graph} to examine the API sequence and classify codes. Researchers also attempted to employ static analysis features for hybrid analysis \cite{shijo2015integrated,islam2013classification}. However, the simplicity of early models and the small amount of training data greatly constrained both the generalization and detection performance.

	\subsection{NLP-based Methods}
	With the significant success of deep learning technology in text classification tasks, researchers are attempting to apply text classification methods to malware dynamic analysis, given that the encoded text bears a high resemblance to API sequences.

Pascanu \textit{et al.} \cite{pascanu2015malware} were the first to suggest the use of Recurrent Neural Networks (RNN) and Echo State Networks (ESN) for dynamic malware detection, pioneering the application of NLP models in dynamic analysis. Nonetheless, RNN demonstrated some degree of gradient vanishing and gradient explosion. To mitigate the issues of the RNN model, researchers \cite{athiwaratkun2017malware,maniath2017deep,catak2020deep,yuan2020character,dang2021malware} utilized the LSTM and GRU models to design dynamic detection models building on Pascanu's work. Apart from RNN-based models, CNN-based \cite{qin2020api,kim2014convolutional} models and CNN+RNN combination models \cite{li2022novel} have also shown good performance. 

Although contemporary methods have achieved solid detection performance, they generally lack robustness and generalization, which are verified in Section~\ref{Representation} and Section~\ref{Domain}.
	
	\subsection{Transformer-based Methods}
	The Transformer \cite{vaswani2017attention} is a deep learning model based on the attention mechanism, first employed in machine translation tasks. The fundamental structure of the Transformer comprises an encoder and a decoder. The encoder is tasked with understanding the input data, while the decoder is used to generate the output results.

Given the success of the Transformer architecture in the NLP field, researchers have sought to transfer this architecture to malware detection models. The Transformer is adept at handling long sequences and can be trained using parallel computing. Moreover, it calculates the attention between each API call to identify key API calls and categorizes the API sequences based on this attention.

Numerous LLMs \cite{devlin2018bert,radford2019language,brown2020language,raffel2020exploring,dai2019transformer,yang2019xlnet} have been proposed, building upon the Transformer model. Some of these models have been applied to malware detection methods. Some research \cite{rahali2021malbert,demirci2022static,rahali2023malbertv2} directly utilize the LLM framework to construct detection models, while others \cite{xu2021malbert,ferrag2023revolutionizing} refer to the pre-training methods of LLMs, obtain the pre-training weights of the model through self-supervised tasks, and then fine-tune it with specific datasets.

It is important to note that the API sequence and text sequence bear somewhat dissimilarities in terms of statistical characteristics and vocabulary set. Furthermore, the local features of the API sequence are typically more significant than its global features. The weak contextual relevance may affect the application of the Transformer module in the malware detection model.
	
		\begin{figure*}[htbp]
		\centering
		\includegraphics[width=1\linewidth]{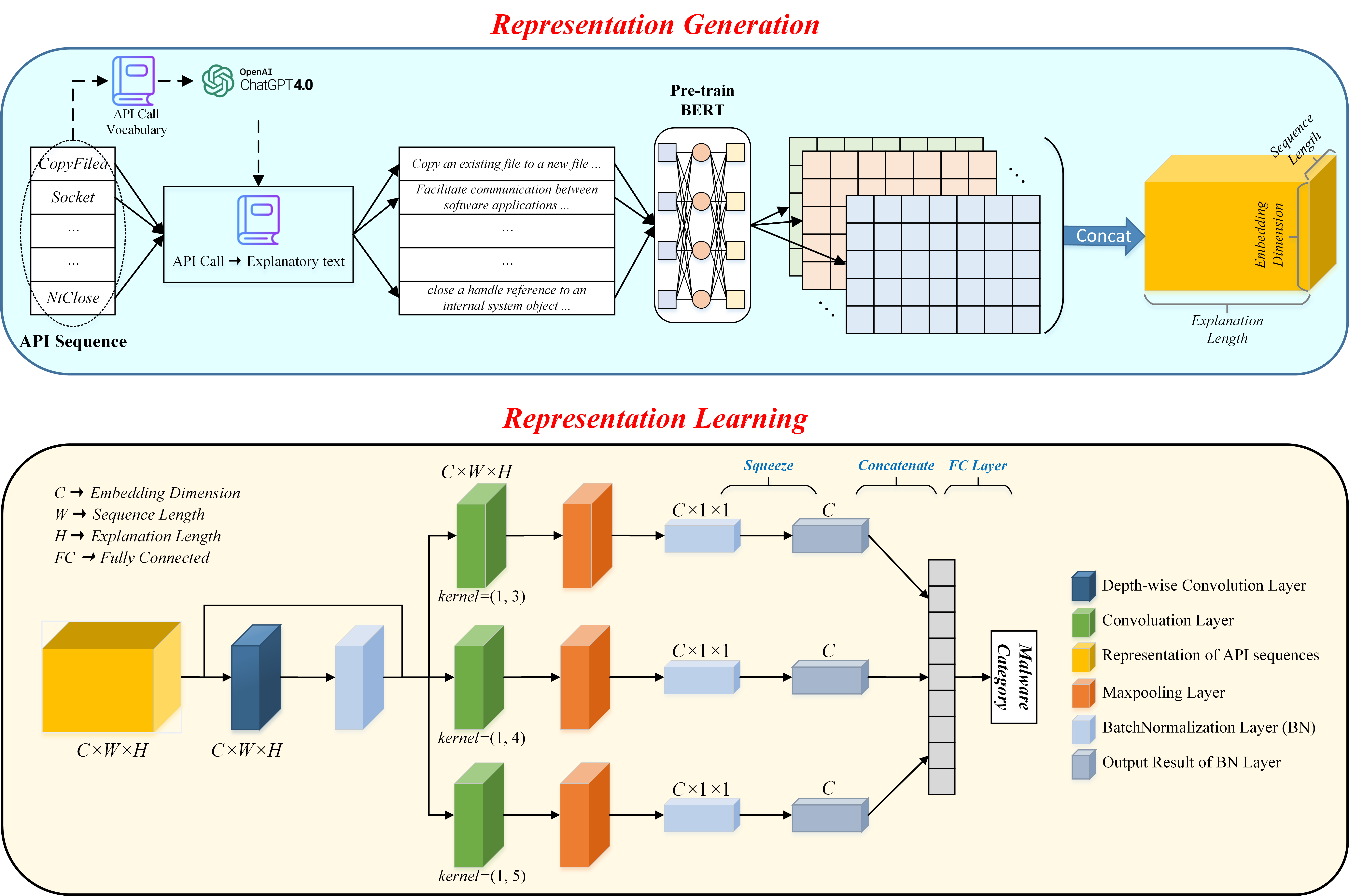}
		\caption{The pipeline of the proposed model is divided into two main modules: Representation Generation and Representation Learning. In the Representation Generation stage, explanatory text for API calls is generated using \textit{GPT-4}. Following this, Bert is used to generate embeddings for the explanatory text, thereby generating the representation of the API sequence. In the Representation Learning phase, a multi-layer convolutional neural network is utilized to extract and subsequently learn feature information from the representation. A fully connected layer is ultimately used to connect to each malware category.}
		\label{Structure}
	\end{figure*}

	\section{Methodology}
	
In this study, \textit{GPT-4} is utilized to produce explanatory text for each API call within the API sequence. Given its training on a large-scale corpus, \textit{GPT-4} can rephrase and summarize the knowledge associated with API calls via prompt engineering. The prompt texts can guide \textit{GPT-4} to generate high-quality explanatory text. Following this, BERT, a large language pre-training model, is employed to generate representations for this explanatory text, which are then concatenated to represent the entire API sequence. The deep neural network is subsequently deployed to extract features from these representations for learning automatically. Finally, the model is connected to various malware code categories through a fully connected layer with a softmax function. The overall architecture of the proposed model is illustrated in Figure~\ref{Structure}.

	\subsection{Representation Generation}
	To generate a representation of the API sequence, we need to produce the explanatory text for each API call in the sequence. For a more detailed depiction of this process, we define a mapping relationship, $Prompt$, wherein we create a sentence for the description and explanation of each API call. We define an API sequence $ s $ with the length of $ n $ as $ s=[\alpha_1,...,\alpha_n] $, where $ \alpha_i $ signifies a single API call. Through prompt engineering, each API call generates descriptive text, and we denote this descriptive text as $ e_i $.

	\begin{equation}
		\begin{split}
			[e_1,...,e_n]=[Prompt(\alpha_1),...,Prompt(\alpha_n)]
		\end{split}.
		\label{eqemb}
	\end{equation}

	However, this method consumes a significant amount of computational resources. Consequently, we construct a vocabulary for the API sequence and generate corresponding explanations for each API call within the vocabulary. 
Subsequently, the explanation text for each API call can be located in the vocabulary, thereby facilitating the reuse of explanation text and significantly reducing computational demands. 

Next, we segment the explanatory text using the $ WordPiece $ segmentation method, as shown in Eq.(~\ref{WordE}).

\begin{equation}
	\begin{split}
		{\tt{[CLS]}},\omega_1,...,\omega_m,{\tt{[SEP]}} =WordPiece(e)
	\end{split}.
	\label{WordE}
\end{equation}	

It decomposes a word into multiple subwords or characters, proving more effective than the space-based method when handling unknown words, rare words, and complex words. 
We segment the explanatory text and tokenize it to obtain a sequence of tokens, then adjust the sequences to the same length $m$, by truncating the exceeding tokens and padding the shortened sequences with the special token {\tt{[PAD]}}. Finally, we incorporate the {\tt{[CLS]}} and {\tt{[SEP]}} special tokens at the beginning and end of the sequence respectively. A mapping relationship $Embed$ is defined as Eq.(~\ref{EmbedE}), generating the vectors that represent each token in the sequence.

	\begin{equation}
		\begin{split}
			\textbf{e}&=Embed([{\tt{[CLS]}},\omega_1,...,\omega_m,{\tt{[SEP]}}]) \\
			&=[v_{CLS},v_1,...,v_m,v_{SEP}].
		\end{split}
		\label{EmbedE}
	\end{equation}
	
 A vector representation, denoted as $ v_j (1 \leq j \leq m) $, is generated for every token $ \omega_k $ within the tokens sequence. The pre-trained BERT is utilized to represent each token, and the dimension of its embedding layer is 768, thus $v_j \in \mathbb{R}^{768}$. This further creates a representation matrix $\textbf{e}_k (1 \leq k \leq n) \in \mathbb{R}^{(m+2)\times 768}$ that corresponds to one API call. Upon obtaining the representation of each API call, a concatenation, denoted as $ Concat $, is performed on the representations of each API call $\textbf{e}_k $ in the API sequence. This process results in the representation tensor \textit{\textbf{E} } $\in \mathbb{R}^{n \times (m+2) \times 768}$ of the API sequence, as shown in Eq.(~\ref{ConcatE}).

	\begin{equation}
		\begin{split}
			\textit{\textbf{E}}=Concat[\textbf{e}_1,...,\textbf{e}_n]
		\end{split}.
		\label{ConcatE}
	\end{equation}
	
	\subsection{Representation Learning}
	The input for previous model methods is a two-dimensional API sequence representation matrix, but the proposed generated representation belongs to a three-dimensional tensor. Consequently, there is a need to design a network architecture capable of accepting a three-dimensional tensor as input, and learning from the representation.

First, to adjust the representation, a depth-wise convolution is performed. The obtained representation is derived from the representation of natural text, which differs somewhat from the API sequence representation. Each embedded channel corresponds to a representation matrix, with each element in the representation matrix having a contextual correlation among the surrounding elements. Specifically, the vertical contextual correlation stems from the explanation text, and the horizontal contextual correlation comes from API sequences. The design of a module for representation adjustments and capturing semantic information is therefore necessary. The trained module can improve the adjustment of the natural text representation for better reflection of API calls and can also capture semantic association information among the surrounding elements.

Considering that the representation of each dimension reflects the specific characteristics of the data and that the correlation of representations across the dimensions is not strong, we employ a per-layer convolutional network to fine-tune each dimension's representation. Additionally, per-layer convolution can capture the correlation and exceptional features of local data.

Unlike natural text sequences, API sequences exhibit significant local features. Therefore, after adjusting the representation, two-dimensional convolution blocks with varying kernel sizes are used to generate respective feature maps. Max pooling and batch normalization operations are then performed on the feature maps. The max pooling operation can select the maximum value from each feature map, allowing the model to capture the important features in each map. It should be considered that the dimensions of the feature map have different statistical distributions, resulting in significant variations in max pooling results. To standardize the results of max pooling, a batch normalization layer is utilized. Finally, the results are concatenated and each classification category is connected through a fully connected layer containing a softmax function.
	
	\begin{figure*}[htbp]
		\centering
		\includegraphics[width=1\linewidth]{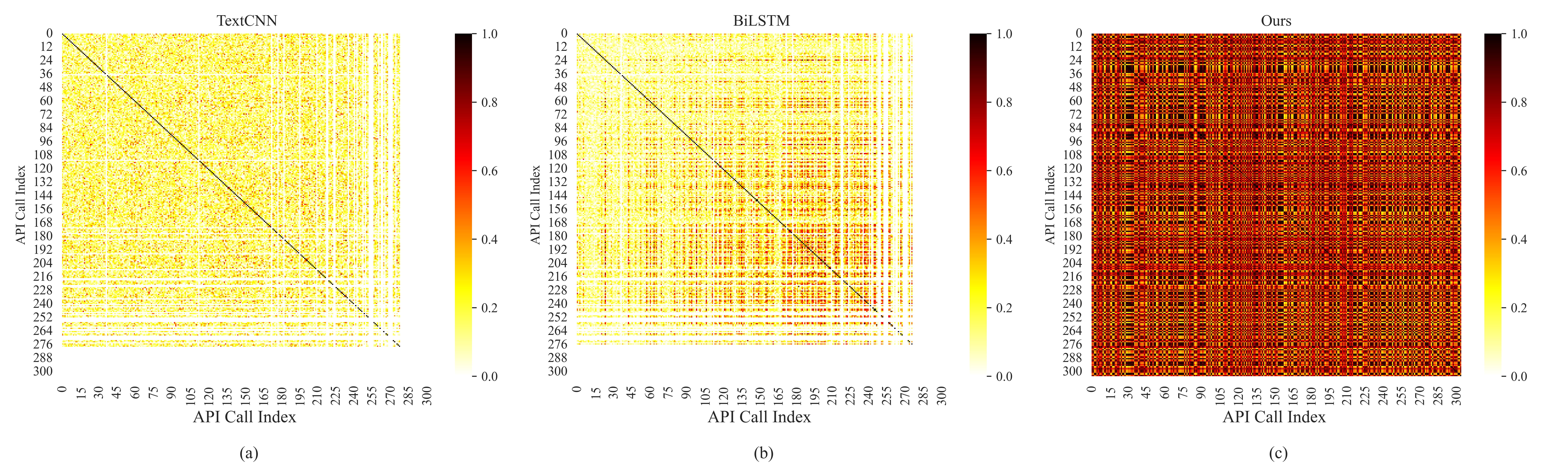}
		\caption{Heat map illustration of the cosine similarity among API call representations produced using different representation techniques, using \textit{Aliyun} as the training dataset. The representations correlation created by the TextCNN \textbf{(a)} and BiLSTM \textbf{(b)} models have many zero values, while the representations correlation derived via the proposed method \textbf{(c)} are more closely related.}
		\label{Cosine}
	\end{figure*}
	
	\subsection{Analysis of Representation Quality}
	\label{QualityS}

The method of representation maps discrete API calls to fixed-size continuous vectors. This method facilitates the calculation of the correlation between each API call through these vectors. For example, vectors corresponding to API calls with similar implications are closer in vector space. Therefore, representation vectors, once trained with datasets or other methods, are compelled to learn and mirror semantic associations between API calls more effectively. This enhances the vectors' ability to deliver a higher quality representation of API calls. Learning based on these high-quality representations, subsequent models can further improve the learning capacity. It is clear that the quality of semantic association in API call representation greatly influences detection performance.

To evaluate the semantic relationships of API calls under different models, we calculate the cosine similarity of API call representation produced by TextCNN, BiLSTM, and the proposed model. The API call representations of TextCNN and BiLSTM are vectors; conversely, the API call representation of the proposed method is a matrix, represented by \textbf{A} and \textbf{B} respectively. The corresponding similarity calculation formula is provided in Eq.(~\ref{CosA}). Since there is no negative correlation between the API call representation in the proposed method, to better showcase the differences in representation effects between our method and the previous methods, we utilize the absolute value of cosine similarity as the measure of representation similarity.

	\begin{equation}
		\begin{split}
			Cosine(\textbf{A},\textbf{B}) = \frac{| \sum_{j=1}^{n}(\textbf{A}_{ij} * \textbf{B}_{ij}) |}{\sqrt{\sum_{j=1}^{n}(\textbf{A}_{ij}^2)} * \sqrt{\sum_{j=1}^{n}(\textbf{B}_{ij}^2)}}
		\end{split}.
		\label{CosA}
	\end{equation}
	
The heat map of API call representation association trained on the \textit{Aliyun} dataset is depicted in Figure~\ref{Cosine}. For the API call representation generated by TextCNN and BiLSTM, there are approximately 18\% of API calls that have almost no correlation with other API calls. One reason for this is that the quantity of training datasets is limited, hence the representation of some API calls cannot be learned. Additionally, after dividing the entire dataset into training and testing datasets, about 15\% types of API calls are absent in the training dataset. Consequently, some correlations of API call representation are not learned during the training process, which results in poor generalization performance.

API call representations are produced by the prompt text of \textit{GPT-4}. So even if certain API calls are absent in the training dataset, the proposed method can generate a representation for these calls and calculate semantic correlations with other API calls. This method is able to generate representations for all API calls, facilitating the calculation of similarity between any two API calls. Hence, the similarity matrix generated by this method is denser and contains more information compared to the previous two methods.

The quality of representation is also a key criterion for evaluating representation generation methods. In the case of API calls, if API calls have similar meanings, then their cosine similarity will be higher. To measure the difference in representational quality between the proposed method and the previous methods, we carry out an analysis of two cases.

		\begin{figure*}[htbp]
	\centering
	\includegraphics[width=0.96\linewidth]{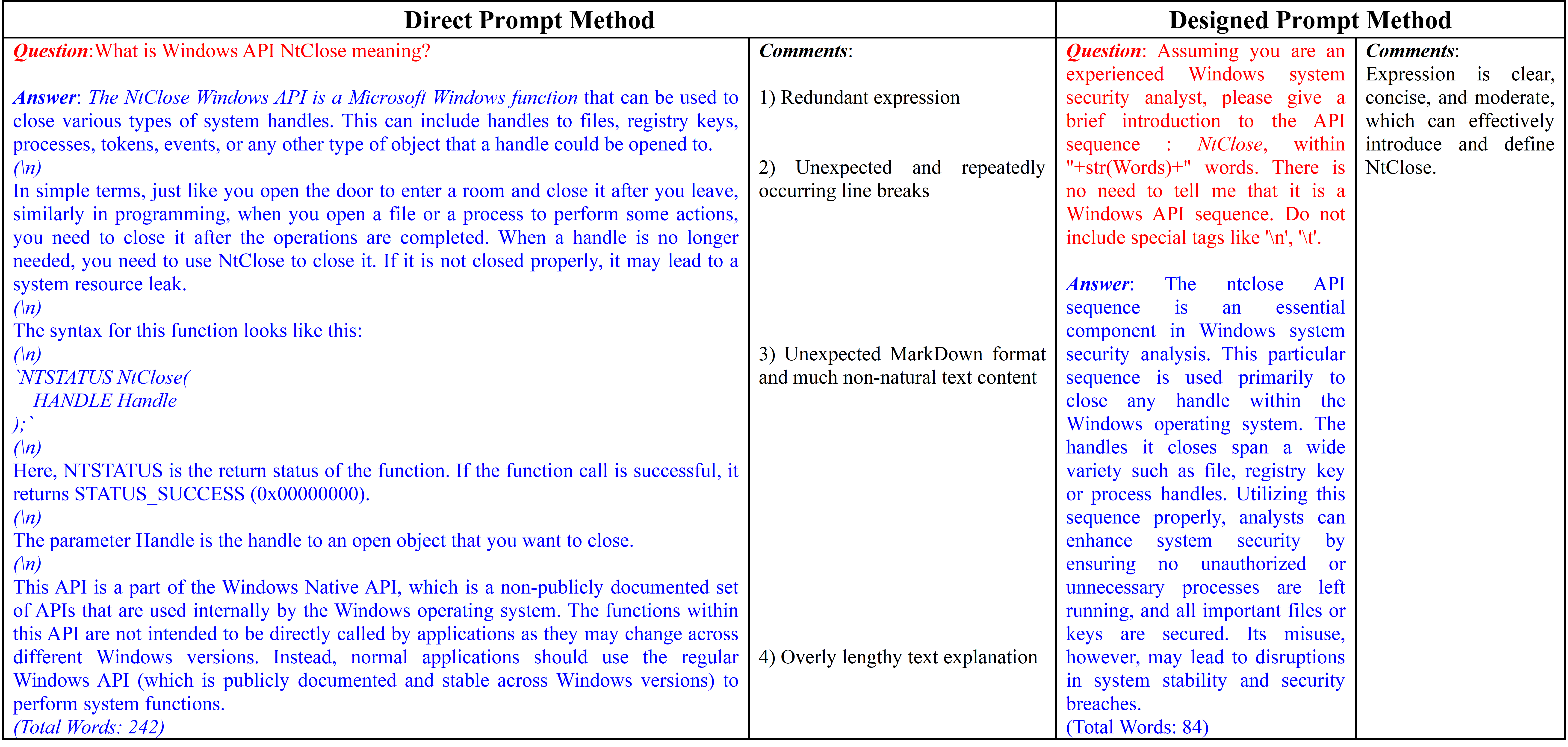}
	\caption{The Comparison of the output content generated by the \textit{GPT-4} model using both the direct prompt method and the designed prompt method. The red text represents the prompt text, while the blue text represents the content output generated by \textit{GPT-4}. The comments on this generated content are indicated in black text. For the answer of the direct prompt method, any text highlighted in \textit{Italics} exhibits problematic content. The specific issues related to this answer are subsequently outlined in the comments provided to the right of these highlighted contents.}
	\label{Prompt}
\end{figure*}

 \noindent\textbf{Case 1: wide and narrow characters.} Consider a pair of API calls, {\tt{HttpSendRequestW}} and {\tt{HttpSendRequestA}}, as an example. The only difference between their names is the final letter. These API calls are two different versions of the same function, with many functions in the Windows API having two versions each dealing with Unicode and ANSI. One version, ending with ``A", deals with narrow characters (ANSI), while the other version, ending with ``W", manages wide characters (Unicode). As such, the meanings of these two API calls are virtually identical, and their cosine similarity is close to one. Previous models learned semantic associations among API calls during the training procedure. However, they struggled to learn correlations under conditions of low data quality or a low occurrence of specific API calls in the datasets. Our method generates explanatory texts for the API calls leading to high similarity in the explanatory texts when their meanings are alike.

 \noindent\textbf{Case 2: semantic chain analysis.} Li \textit{et al.} \cite{li2022novel} proposed a semantic chain method. This method generates four attributes of an API call based on the API call name. These attributes are $ action $, $ object $, $ class $, and $ category $, collectively forming the semantic chain of the API call. If two API calls share the same semantic chain, their meanings are similar and their cosine distance similarities approach one. The case where the four attributes are identical corresponds to Case 1. Considering the strong distinguishing capacity of the $ object $ attribute, we examine a pair of API calls that share identical \textit{action}, \textit{class}, and \textit{category} attributes. The proposed method effectively captures the relationships of API calls with similar semantic chains, posing a challenge for prior representation methods.

Detailed illustrations of the aforementioned two cases, as well as a comparison between the performance of the proposed model and mainstream models on these cases, are provided in Section~\ref{quality}.

To further assess the representation quality of the model, we train and test the models with two different datasets, a process known as cross-database experiments. When the similarity of the API call vocabulary is high, the representation formed by the training datasets must adapt to the representation of the test datasets, a process referred to as \textit{Representation Adaptation}. Conversely, when the vocabularic similarity is low, many API calls in the testing dataset would not have been encountered during training. This lack of significant feature representation poses challenges when testing on the testing dataset. We treat the API call vocabularies of the training and testing datasets as two minimally overlapping domains and apply the domain knowledge learned from the training dataset to the testing domain in a process termed \textit{Domain Adaptation}. Domain adaptation exerts greater demands on generalization capabilities compared to representation adaptation, making it a considerable challenge for dynamic analysis models. Detailed accounts of the corresponding experiments and analyses are provided in Section~\ref{Representation} and Section~\ref{Domain}.

	\subsection{Design of Prompt Texts}

Once the model achieves a certain scale, its performance significantly improves, demonstrating strong capabilities, such as language comprehension, generation ability, logical reasoning, and so forth. Therefore, designing prompt texts that enable these large-scale models to exhibit such powerful capabilities is worth exploring.
Wei \textit{et al.}\cite{wei2022chain} suggested an enhanced strategy for generating prompt text, Chain of Thing (CoT). By providing auxiliary prompts for intermediate reasoning steps, CoT allows large models to tackle more complex problems.
	
\begin{figure}[!t]
		\centering
		\includegraphics[width=1\linewidth]{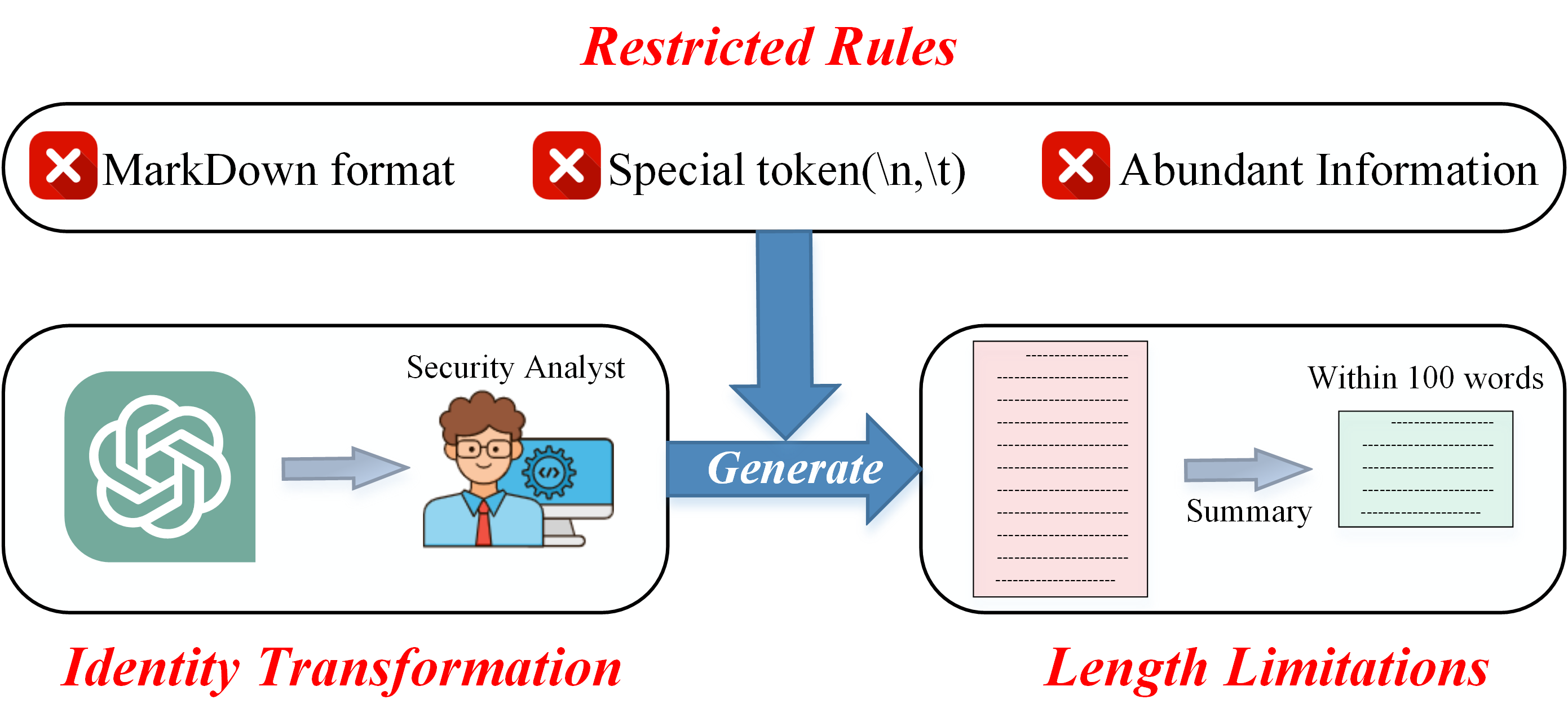}
		\caption{The design process entails the creation of a prompt text, which guides \textit{GPT-4} in generating explanatory text of a higher quality.}
		\label{CoT}
	\end{figure}

In this paper, the representation is created by the \textit{GPT-4}. Even though the \textit{GPT-4} is not required to execute complex reasoning, it must paraphrase its learned knowledge taking into account specific requirements. In this process, the prompt text directly influences the quality of the prompt content. Consequently, we take into account the following rules for designing prompt text, and the comprehensive design process of the prompt text is depicted in Figure.~\ref{CoT}.

\begin{itemize}
\item \noindent\textbf{Identity Transformation.} Yang \textit{et al.}\cite{yang2023dawn} demonstrated that hypothesizing specific identities and operating environments to the \textit{GPT-4} boosts its level of expression and reasoning, and thereby generates higher-quality prompt content. Therefore, we treat \textit{GPT-4} like an experienced software security analyst capable of carrying out the malware analysis task with high quality.

\item \noindent\textbf{Restricted Rules.} We explicitly instruct \textit{GPT-4} not to generate redundant content such as ``XXX is a Windows API sequence". This attribute is a typical characteristic of the API calls and cannot be used to differentiate them. Moreover, \textit{GPT-4} is required to present the generated content in a natural text form, without adding special symbols (\textit{e.g.}, ``\textbackslash n", ``\textbackslash t") or presenting content in unusual formats (\textit{e.g.}, Markdown format).

\item \noindent\textbf{Length Limitation.} We have to accomplish \textit{WordPiece} tokenization on the generated text and process the token sequence to a fixed length to ensure that the representations produced from each text have the same form. Thus, we explicitly demand that the text created by \textit{GPT-4} is restricted to 100 words. Text that is too lengthy will substantially increase the time consumption and computation space required.
\end{itemize}

Finally, we input both the direct prompt text and the designed prompt text into the \textit{GPT-4}, respectively. The comparison of the content generated by \textit{GPT-4} is illustrated in Figure~\ref{Prompt}. Clearly, if we guide \textit{GPT-4} with the designed prompt text, the content created is of superior quality, which finally improves the detection performance.

\section{Experiments of Detection Models}

Five benchmark datasets are employed to evaluate the performance of the proposed model. The selection of two high-performance models (TextCNN, BiLSTM) is based on detection accuracy for further analysis. To assess the generalization performance of the proposed model, five datasets are classified into two groups, according to the association of the API call vocabulary. Concurrently, representation adaptation experiments are trained and tested within the same group, while domain adaptation representations are tested across different groups.
\subsection{Experiment Settings}
\noindent\textbf{Implementation Details}. All experiments in this paper are carried out on Ubuntu 20.04, utilizing an RTX 4090 GPU and 24 GB of memory. Python 3.9 and Pytorch 2.0 are used to construct the experimental model. Considering the GPU memory capacity limitations, the truncation length of the API sequence is set at 100, and the embedded token sequence of the explanatory text is set at 102 (including the initial {\tt{[CLS]}} and final {\tt{[SEP]}} token). The batch size is set at 8, and the learning rate is set at 0.001, with the Adam optimizer utilized.

\noindent\textbf{Compared Models.} The models used for comparison can be mainly grouped into several categories: RNN-based networks \cite{dang2021malware,yang2023dawn,catak2020deep,zhang2020dynamic}, CNN-based networks \cite{qin2020api}, CNN+RNN-based networks \cite{kolosnjaji2016deep,li2022novel,zhang2023dynamic}, and Transformer-based networks \cite{demirkiran2022ensemble,xu2021malbert}.

\noindent\textbf{Datasets.} To validate the effectiveness of the proposed model, it is trained or tested using five benchmark datasets of malware dynamic API call sequences: \textit{Aliyun} \cite{dataset:mal_sandbox}, \textit{Catak} \cite{catak2020deep}, \textit{GraphMal} \cite{oliveira2019behavioral}, \textit{VirusShare} \cite{Virus}, and \textit{VirusSample} \cite{Virus}. Based on the similarity of their respective API call vocabularies, the five datasets are divided into two sets:
	$$
	\mathcal{D}_{base}=\{Aliyun,Catak,GraphMal\},
	$$
	$$
	\mathcal{D}_{large}=\{VirusSample,VirusShare\}.
	$$
The number of vocabulary in the datasets of $\mathcal{D}_{large}$ is significantly higher than in that of $\mathcal{D}_{base}$. Besides, the $\mathcal{D}_{large}$ is more complex and contains abnormal contents. The descriptive statistical features of these five datasets are displayed in Table ~\ref{datasets}.

With regards to the \textit{Aliyun} dataset, the proportion of malicious sequences is quite low, which could potentially impede the recall rate of unrecognized malicious sequences. As a solution, the \textit{Aliyun} and \textit{Catak} datasets are merged to produce an expanded \textit{Aliyun+Catak} dataset. The benign sequences originating from the \textit{Aliyun} dataset form the base of benign sequences in this composite dataset. Conversely, the malicious sequences from all categories within the \textit{Aliyun} dataset, along with all sequences from the \textit{Catak} dataset, constitute the malicious sequences in this combined dataset. By increasing the proportion of malicious sequences in the dataset, it is expected to enhance the model's recall rate for unknown malicious sequences.
	
	\begin{table*}[]
		\caption{Statistics of the 5 benchmark datasets}
		\adjustbox{center}{
			\resizebox{1.0\textwidth}{!}{
				\renewcommand{\arraystretch}{1.3}	
				\begin{tabular}{rccccc}
					\Xhline{3\arrayrulewidth}
					\textbf{Dataset} & \textbf{\begin{tabular}[c]{@{}c@{}}Proportion \\ of benign\end{tabular}} & \textbf{\begin{tabular}[c]{@{}c@{}}Proportion \\ of malicious\end{tabular}} & \textbf{\begin{tabular}[c]{@{}c@{}}Samples \\ Amount\end{tabular}} & \textbf{\begin{tabular}[c]{@{}c@{}}Vocabulary Size \\ of API Call\end{tabular}} & \textbf{Category Distribution}           \\ \hline
					\textit{Aliyun}           & 64.15\%                                                                  & 35.85\%                                                                     & 13887                                                              & 301                                                                             & 1 kind of  benign and 7 kinds of malware \\
					\textit{Catak}            & 0\%                                                                      & 100\%                                                                       & 7107                                                               & 281                                                                             & 8 kinds of malware                       \\
					\textit{Aliyun+Catak}     & 23.71\%                                                                  & 76.29\%                                                                     & 20994                                                              & 304                                                                             & 1 kind of  benign and 1 kind of  malware \\
					\textit{GraphMal}         & 2.46\%                                                                   & 97.54\%                                                                     & 43876                                                              & 304                                                                             & 1 kind of  benign and 1 kind of  malware \\
					\textit{VirusSample}      & 0\%                                                                      & 100\%                                                                       & 9795                                                               & 7964                                                                            & all are malwares                         \\
					\textit{VirusShare}       & 0\%                                                                      & 100\%                                                                       & 14616                                                              & 23229                                                                           & all are malwares                        \\ \Xhline{3\arrayrulewidth}
				\end{tabular}
			}
		}
		\label{datasets}
	\end{table*}
	
	\subsection{Statistical Properties Analysis of Datasets}
	
For a more intuitive understanding of each dataset, the statistical characteristics of the datasets used in the experiments are depicted in Table ~\ref{datasets}. We construct an API call vocabulary for each dataset and calculate the similarity between each API call vocabulary. IoU is adopted as the similarity measurement criteria, and its formula is provided in Eq.(~\ref{IoUE}).
\begin{equation}
\begin{split}
IoU(\mathcal{S}_1,\mathcal{S}_2)=\frac{|\mathcal{S}_1 \cap \mathcal{S}_2|}{|\mathcal{S}_1 \cup \mathcal{S}_2|}
\end{split}, 
\label{IoUE}
\end{equation}
where $\mathcal{S}_1 $ and $ \mathcal{S}_2 $ denote the API call vocabulary for different datasets, respectively. The similarity among the API call vocabularies consistently increases as the value of IoU rises.

\begin{figure}[htbp]
	\centering
	\includegraphics[width=0.95\linewidth]{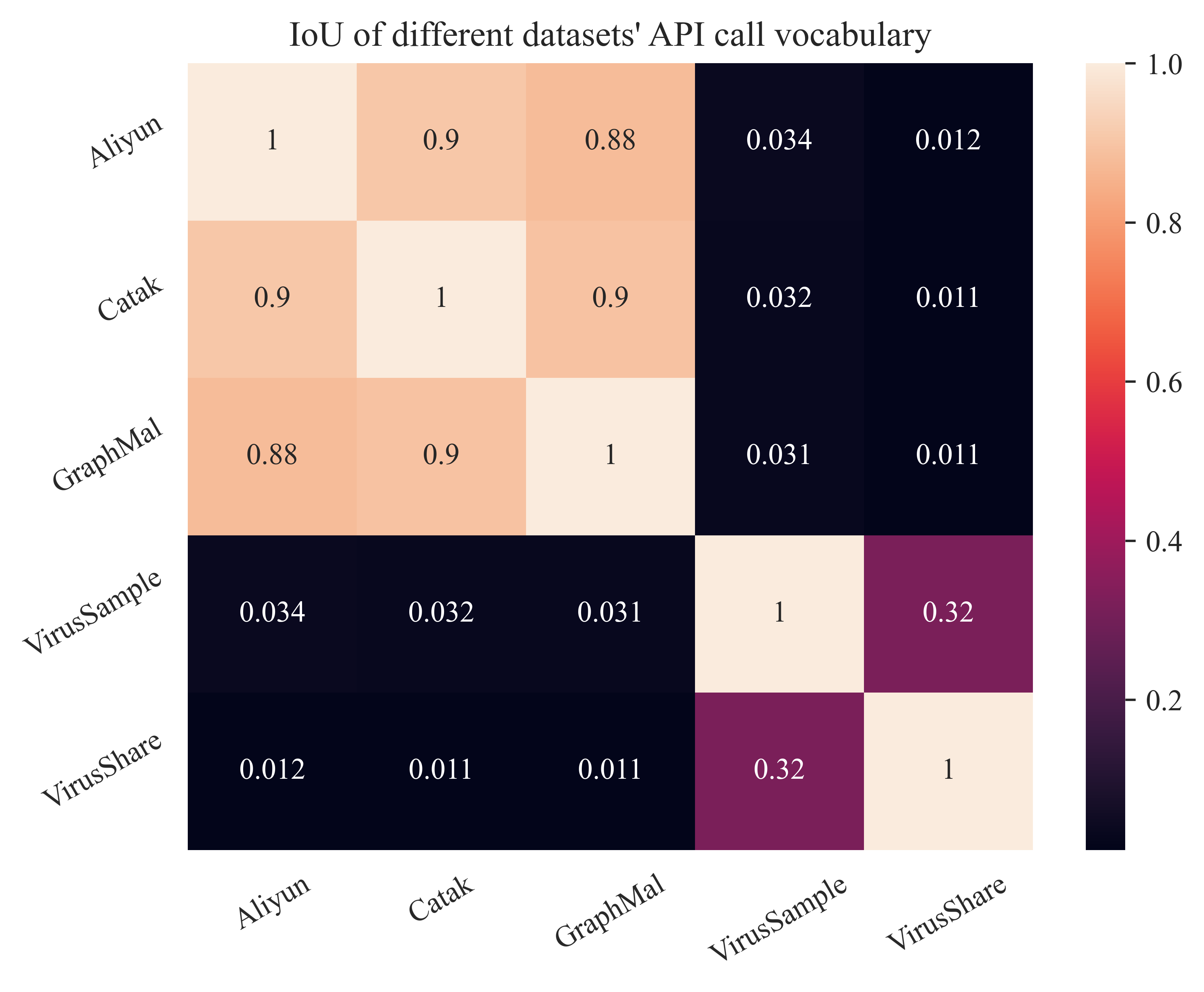}
	\caption{The heatmap represents the Intersection over Union (IoU) values of API call vocabularies contained in different datasets.}
	\label{IoU}
\end{figure}

The IoU values for the API call vocabulary of each dataset are depicted in Figure~\ref{IoU}. There is a significant level of similarity among the \textit{Aliyun}, \textit{Catak}, and \textit{GraphMal} datasets, whereas these three datasets exhibit markedly low resemblance to the \textit{VirusSample} and \textit{VirusShare} datasets. This disparity is attributed to the dissimilar methods of dynamic feature extraction. Predominantly, \textit{Aliyun}, \textit{Catak}, and \textit{GraphMal} record high-level API calls. By contrast, the \textit{VirusSample} and \textit{VirusShare} datasets have a more complex structure. They not only include high-level API calls but are also rich in low-level API calls, with some even presenting anomalies. This results in a comprehensive and considerably greater API call vocabulary for these datasets than for the aforementioned datasets in $\mathcal{D}_{base}$.

	\subsection{Performance of the Proposed Model}
	
		\begin{table*}[]
		\caption{Comparison of the Performance of Different Detection Models.}
		\adjustbox{center}{
			\resizebox{1.0\textwidth}{!}{
				\renewcommand{\arraystretch}{1.3}	
				\begin{tabular}{||r|cccclcccc}
					\Xhline{3\arrayrulewidth}
					&                                     & \multicolumn{3}{c}{\textbf{\textit{Aliyun}}}&                       & \textbf{\textit{Catak}} &     & \multicolumn{2}{c}{\textbf{\textit{GraphMal}}}   \\ \cline{3-5} \cline{7-7} \cline{9-10} 
					\multirow{-2}{*}{\textbf{Method}}       & \multirow{-2}{*}{\textbf{Type}}     & \textbf{Multi(ACC)} & \textbf{Binary(ACC)} & \textbf{AUC} & & \textbf{Multi(ACC)} & & \textbf{Binary(ACC)}  & \textbf{AUC}  \\ \hline
					BiLSTM\cite{dang2021malware}                                  &                                     & 82.65\%             & 93.38\%              & 0.9813       & & 49.51\%             & & 99.38\%               & 0.9887        \\
					BiGRU\cite{yuan2020character}                                   &                                     & 81.43\%             & 93.52\%              & 0.9825       & & 49.65\%             & & 99.45\%               & 0.9766        \\
					CatakNet\cite{catak2020deep}                                &                                     & 82.22\%             & 93.45\%              & 0.9791       & & 49.09\%             & & 98.81\%               & 0.9733        \\
					ZhangNet\cite{zhang2020dynamic}                                & \multirow{-4}{*}{RNN-based}         & 77.75\%             & 89.85\%              & 0.9512       & & 40.79\%             & & 97.54\%               & /             \\ \hdashline
					Kolosnjaji\cite{kolosnjaji2016deep}                              &                                     & 81.57\%             & 93.38\%              & 0.9751       & & 45.15\%             & & 99.32\%               & 0.9895        \\
					LiNet\cite{li2022novel}                                   &                                     & 79.12\%             & 93.74\%              & 0.9522       & & 48.10\%             & & 97.54\%               & 0.6010         \\
					Mal-ASSF\cite{zhang2023dynamic}                                & \multirow{-3}{*}{CNN+RNN-based}     & 82.36\%             & 93.81\%              & 0.9815       & & 48.66\%             & & 98.98\%               & 0.9390         \\ \hdashline
					TextCNN\cite{qin2020api}                                 & CNN-based                           & 83.44\%             & 94.53\%              & 0.9847       & & 47.96\%             & & 99.36\%               & 0.9936        \\ \hdashline
					Transformer\cite{demirkiran2022ensemble}                             &                                     & 75.95\%             & 91.07\%              & 0.9643       & & 37.83\%             & & 98.59\%               & 0.9216        \\
					MalBert\cite{xu2021malbert}                                 & \multirow{-2}{*}{Transformer-based} & 77.83\%             & 89.99\%              & 0.9579       & & 38.82\%             & & 97.49\%               & 0.5003        \\ \hline
					{Embed3D+CNN} &                                     & 82.29\%             & 94.53\%              & 0.9848       & & 52.32\%             & & 98.97\%               & 0.9905        \\
					{\textbf{\textit{Ours}}} & \multirow{-2}{*}{CNN-based}         & \textbf{85.89\%}            & \textbf{95.61\%}              & \textbf{0.9923}       & & \textbf{62.03\%}            & & \textbf{99.45\%}              & \textbf{0.9976}        \\ \Xhline{3\arrayrulewidth}& & 
				\end{tabular}
			}
		}
		\label{performance}
	\end{table*}

We compare the performance of the proposed model with the SOTA model using three datasets (\textit{Aliyun}, \textit{Catak} and \textit{GraphMal}). To validate the effectiveness of the proposed representation generation method, we carry out an ablation study. Keeping the representation learning module unaltered, we employ the embedding layer to create the representation matrix of the API sequence, then duplicate this matrix to construct the representation tensor. This is designed to match the shape ($\in \mathbb{R}^{100*102*768}$) of the representation produced by the proposed method. This method is denoted as \textit{Embed3D+CNN}.

As shown in Table ~\ref{performance}, the proposed model demonstrates improved detection performance on all three datasets in comparison to SOTA methods. The integration of additional external knowledge during training somewhat enhances the performance of the model. However, the extent of this improvement is not substantial. This method essentially employs textual representation as the API call representation, which results in a small discrepancy between these two types of representation. Although the convolutional neural network (CNN) adjusts the representation to bridge the gap between them, the outcome is limited due to the constraints of training on a finite dataset. Moreover, key API calls appear repeatedly across multiple training iterations, giving other models the advantage of learning their representation and consequently achieving commendable detection results.
                                                                                                      	
	\subsection{Representation Adaptation}
	\label{Representation}
	
	In this experiment, the divergence in API call vocabulary between the training and testing sets is small, thus meaning that most API calls in the testing set already exist in the training set.

The focus of this experiment is to verify the generalization performance of the model, with the results displayed in Table ~\ref{strong}. The representation generated by the proposed method is of higher quality, with denser associations and enhanced stability. As a consequence, the detection performance of our method surpasses others in cross-database experiments, affirming the generalization performance of our model.
Regarding other models, despite their representation associations having many zero values, the key associations have been learned within the training set; thus, these also exhibit a certain degree of detection effect and generalization ability.
In the experiment where \textit{Aliyun} is used as the training set and \textit{GraphMal} as the test set, the number of malware samples in \textit{Aliyun} is fewer, leading to a lower recall rate of malware. Consequently, we introduce malware samples from the \textit{Catak} dataset, utilizing a combined dataset of \textit{Aliyun} and \textit{Catak} for training. Upon validating the trained model on \textit{GraphMal}, both the recall rate of malware and the overall accuracy significantly improve.

	\subsection{Domain Adaptation}
	\label{Domain}
	In this experiment, there is a significant divergence in API call vocabulary between the training set and the testing set. As a result, most API calls in the testing set have not been encountered during the training process. Numerous API calls present in $ \mathcal{D}_{large} $ do not exist in $ \mathcal{D}_{base} $. This disparity introduces significant complications for cross-database experiments. 

However, the proposed method can generate explanatory text and corresponding representation for an unseen API call encountered during the training process. This capability significantly reduces the impact of representation absence on the prediction effect of the model. Despite being trained on $ \mathcal{D}_{base} $, the proposed method exhibits high detection performance on $ \mathcal{D}_{large} $. The recall rate of malware is nearly 100\%.

By contrast, models, except those trained on the \textit{GraphMal} dataset, display virtually no prediction ability on $ \mathcal{D}_{large} $ when trained using $ \mathcal{D}_{base} $. The \textit{GraphMal} dataset consists of 98\% malware samples, creating a prediction bias towards the malware category during testing. Consequently, the recall rate of other methods is able to achieve such a high level.

\section{Experiments of Representation Quality}	
	Using the semantic chain similarity of API calls as a reference, we examine the representation quality of various models. Additionally, we investigate the performance of generating representations using API calls from differing sources. Ultimately, we assess the efficacy of the proposed model in addressing the phenomenon of concept drift.
	
	\subsection{Comparison of Representation Quality}
	\label{quality}
	\begin{table*}[]
		\caption{Comparison of Different Models' Representation Quality. In instances where the API calls share the same meaning but manage different character types.}
		\adjustbox{center}{
			\resizebox{0.7\textwidth}{!}{
				\renewcommand{\arraystretch}{1.3}	
				\begin{tabular}{rcccc}
					\Xhline{3\arrayrulewidth}
					\textbf{API Call 1}  & \textbf{API Call 2}  & {\textbf{Ours}} & \textbf{TextCNN} & \textbf{BiLSTM} \\ \hline
					{\tt{RegQueryValueExW}}     & {\tt{RegQueryValueExA}}     & {\textbf{0.8514}}        & -0.1253          & -0.1455         \\
					{\tt{WSASocketW}}           & {\tt{WSASocketA}}            & {\textbf{0.7055}}        & -0.8257          & -0.4443         \\
					{\tt{SetWindowsHookExW}}    & {\tt{SetWindowsHookExA}}    & {\textbf{0.9478}}        & 0.1864           & -0.0022         \\
					{\tt{DeleteUrlCacheEntryW}} & {\tt{DeleteUrlCacheEntryA}} & {\textbf{0.8507}}        & 0                & 0               \\
					{\tt{HttpOpenRequestW}}     & {\tt{HttpOpenRequestA}}     & {\textbf{0.7520}}         & 0                & 0               \\ \Xhline{3\arrayrulewidth}
				\end{tabular}
			}
		}
		\label{case1}
	\end{table*}
	
	\begin{table*}[]
		\caption{Comparison of the Representation Quality of Different Models. This pertains to instances where the API calls possess identical semantic chains.}
		\adjustbox{center}{
			\resizebox{0.9\textwidth}{!}{
				\renewcommand{\arraystretch}{1.3}	
				\begin{tabular}{rccccccc}
					\Xhline{3\arrayrulewidth}
					&                                       & \multicolumn{3}{c}{\textbf{Semantic Chain}} & {\color[HTML]{FD6864} }                                &                                    &                                   \\ \cline{3-5}
					\multirow{-2}{*}{\textbf{API Call 1}} & \multirow{-2}{*}{\textbf{API Call 2}} & \textit{action}       & \textit{class}        & \textit{category}      & \multirow{-2}{*}{{\textbf{Ours}}} & \multirow{-2}{*}{\textbf{TextCNN}} & \multirow{-2}{*}{\textbf{BiLSTM}} \\ \hline
					{\tt{NtUnloadDriver}}                        &{\tt{LdrUnloadDll}}                          & Update       & system       & Unload        & {\textbf{0.7904}}                          & 0                                  & 0                                 \\
					{\tt{NtOpenDirectoryObject}}                 & {\tt{NtOpenFile}}                            & Update       & file         & Open          & {\textbf{0.8733}}                          & 0.0544                             & -0.1711                           \\
					{\tt{WSASendTo}}                             & {\tt{WSASend}}                               & Update       & network      & Send          & {\textbf{0.7568}}                          & 0                                  & 0                                 \\ \Xhline{3\arrayrulewidth}
				\end{tabular}
			}
		}
		\label{case2}
	\end{table*}
	
	We propose two case studies to measure the representation quality in Section~\ref{QualityS}. The first one is the wide character and narrow character method where the wide and narrow versions of the same function have different API names, yet their representational meanings are remarkably similar. The second method is the representational semantic chain association method. In this case, if the semantic association chains of two API calls are the same, their representational meanings are regarded as similar as well. 
	Some API call examples are analyzed with two case studies, and the results are exhibited in Table~\ref{case1} and Table~\ref{case2} respectively.

\begin{figure}[]
		\centering
		\includegraphics[width=1\linewidth]{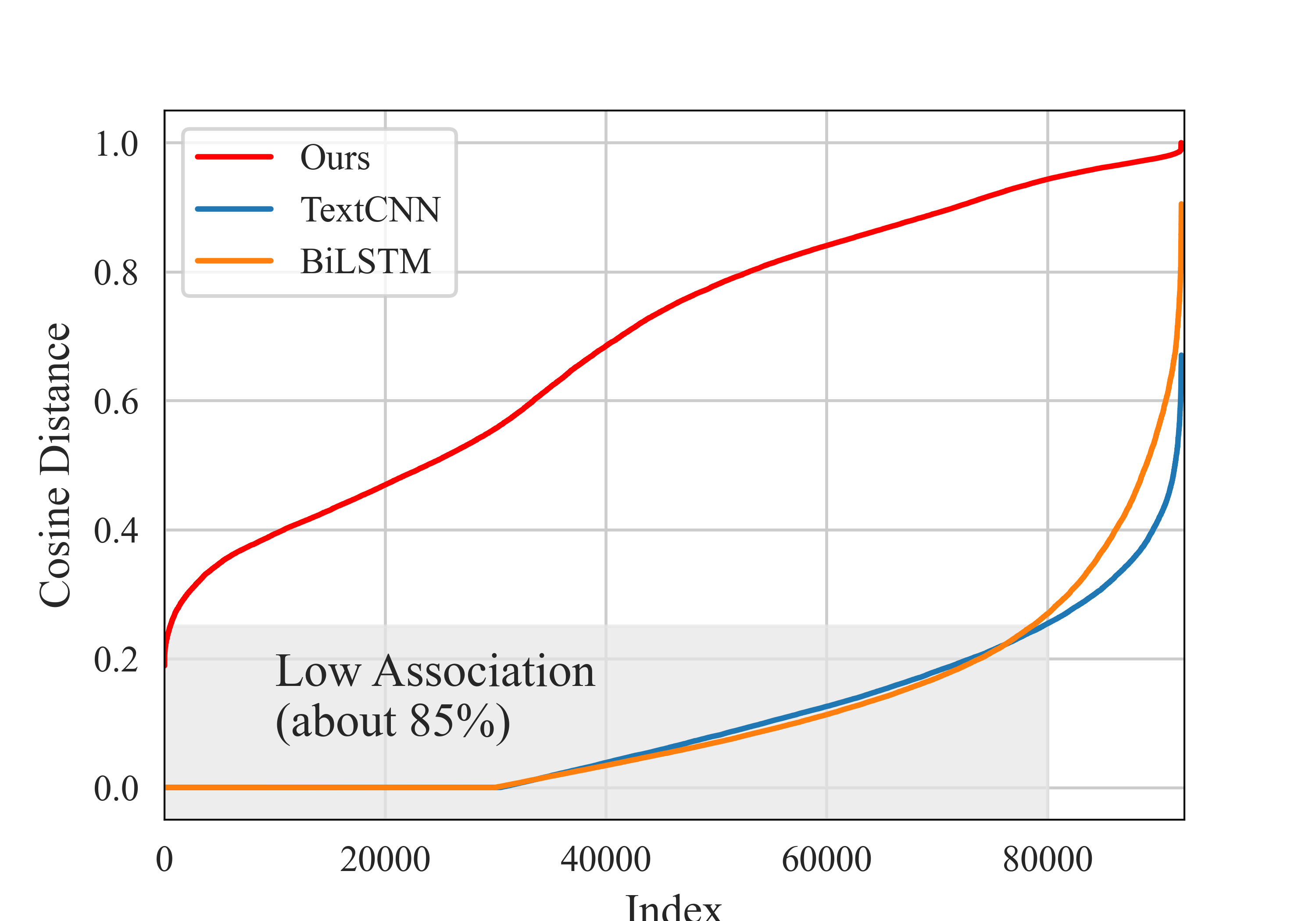}
		\caption{
			The statistical distribution of API call association similarity. The API call association matrix (Figure.~\ref{Cosine}) is first flattened and each element in flattened matrix is arranged in ascending order. When the association falls below 25\%, we infer that there is virtually no association between the two API calls. In the case of TextCNN and BiLSTM models, approximately 85\% of the API call representations lack any association.} 
		\label{association}
	\end{figure}

Our proposed method can generate denser representations and capture the associations between API calls as effectively as possible. Therefore, in both case studies, it accurately generates the association between API call representations. However, in the case of TextCNN and BiLSTM, their representations have to be obtained via dataset training. Hence, the API call association is constrained by the quality of the training datasets. As illustrated in Figure~\ref{association}, approximately 85\% of association degrees in the trained API call representations fall between [-0.25,0.25]. This limitation stems from the poor quality of the dataset, as a result, these methods have difficulty in generating a wider range of API call associations.
	
	\begin{table}[]
		\caption{Comparison of Model Performance in Representation Adaptation Experiments}
		\adjustbox{center}{
			\resizebox{0.5\textwidth}{!}{
				\renewcommand{\arraystretch}{1.3}	
				\begin{tabular}{rrcccc}
					\Xhline{3\arrayrulewidth}
					\textbf{Training}                & \textbf{Testing}         & \textbf{Model}   & \textbf{Precision} & \textbf{Recall} & \textbf{ACC} \\ \hline
					\multirow{3}{*}{\textit{Aliyun}}       & \multirow{3}{*}{\textit{Catak}}    & Ours             & /                  & \textbf{51.48\%}           & /            \\
					&                           & TextCNN          & /                  & 50.44\%           & /            \\
					&                           & BiLSTM           & /                  & 50.96\%           & /            \\ \hdashline
					\multirow{3}{*}{\textit{Aliyun}}       & \multirow{3}{*}{\textit{GraphMal}} & Ours             & \textbf{99.40\%}              & \textbf{30.58\%}           & \textbf{32.10\%}         \\
					&                           & TextCNN          & 99.21\%              & 25.30\%            & 26.94\%        \\
					&                           & BiLSTM           & 98.75\%              & 17.59\%           & 19.40\%         \\ \hdashline
					\multirow{3}{*}{\textit{Aliyun+Catak}} & \multirow{3}{*}{\textit{GraphMal}} & Ours             & 98.31\%              & \textbf{62.07\%}           & \textbf{61.96\%}        \\
					&                           & TextCNN          & 98.24\%              & 54.09\%           & 54.27\%        \\
					&                           & BiLSTM           & \textbf{98.77\%}              & 56.26\%           & 56.65\%        \\ \hdashline
					\multirow{3}{*}{\textit{GraphMal}}     & \multirow{3}{*}{\textit{Aliyun}}   & Ours             & \textbf{73.71\%}              & 91.02\%           & \textbf{72.79\%}       \\
					&                           & TextCNN          & 68.94\%              & 90.93\%           & 67.91\%        \\
					&                           & BiLSTM           & 69.52\%              & \textbf{94.49\%}           & 69.89\%        \\ \hdashline
					\multirow{3}{*}{\textit{GraphMal}}     & \multirow{3}{*}{\textit{Catak}}    & Ours             & /                  & \textbf{99.97\%}           & /            \\
					&                           & TextCNN          & /                  & 85.87\%           & /            \\
					&                           & BiLSTM           & /                  & 77.59\%           & /           \\ \Xhline{3\arrayrulewidth}
				\end{tabular}
			}
		}
		\label{strong}
	\end{table}

	\begin{table}[htb]
		\centering
		\caption{Comparison of Model Performance in Domain Adaptation Experiments}
		\adjustbox{center}{
			\resizebox{0.5\textwidth}{!}{
				\renewcommand{\arraystretch}{1.3}	
				\begin{tabular}{rcccc}
					\Xhline{3\arrayrulewidth}
					\multirow{2}{*}{\textbf{Training}} & \multirow{2}{*}{\textbf{Testing}} & \multicolumn{3}{c}{\textbf{Recall}}                                        \\ \cline{3-5} 
					&                           & Ours    & TextCNN & BiLSTM  \\ \hline
					\textit{GraphMal}               & \textit{VirusSample}               & \textbf{99.15\%} & 98.97\% & 94.97\% \\
					\textit{Aliyun+Catak}           & \textit{VirusSample}               & \textbf{100\%}   & 17.75\% & 51.07\% \\
					\textit{Aliyun}                 & \textit{VirusSample}               & \textbf{99.04\%} & 24.19\% & 54.98\% \\ \hdashline
					\textit{GraphMal}               & \textit{VirusShare}                & 99.90\% & 99.88\% & \textbf{99.97\%} \\
					\textit{Aliyun+Catak}           & \textit{VirusShare}                & \textbf{100\%}   & 37.70\% & 72.94\% \\
					\textit{Aliyun}                 & \textit{VirusShare}                & \textbf{94.53\%} & 24.99\% & 29.19\% \\ \Xhline{3\arrayrulewidth}
				\end{tabular}
			}
		}
		\label{weak}
	\end{table}
	
	\begin{figure*}[!t]
		\centering
		\includegraphics[width=1\linewidth]{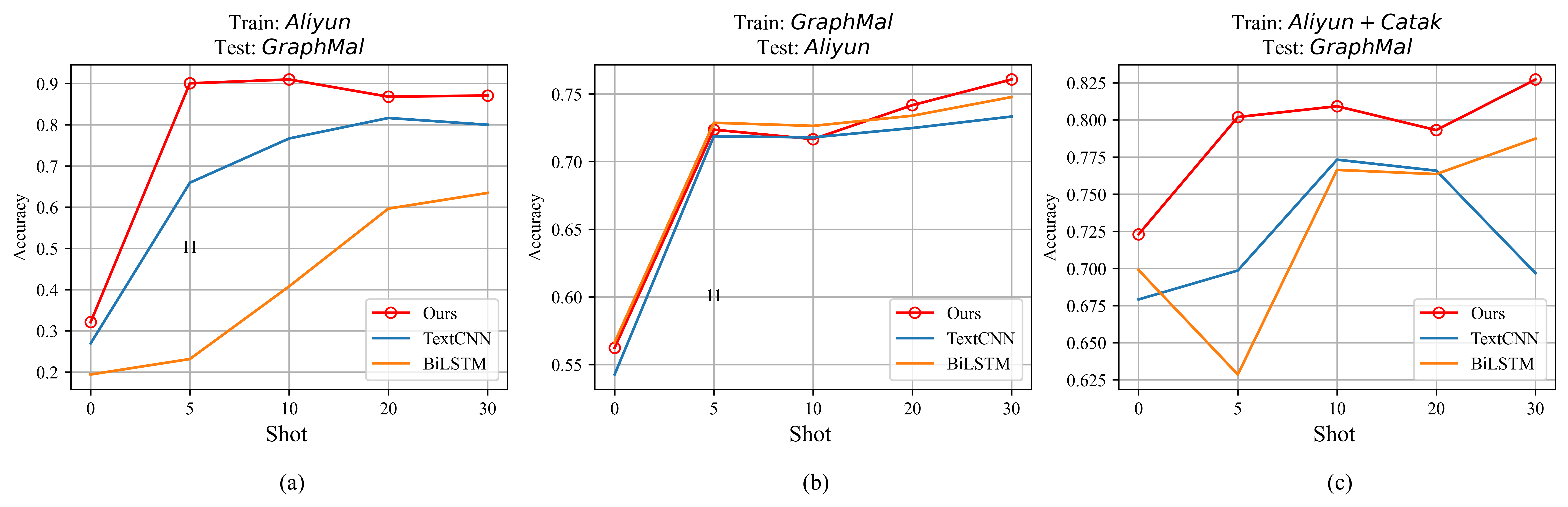}
		\caption{The results of few-shot fine-tuning experiments for different models. \textbf{(a)}, \textbf{(b)} and \textbf{(c)} are trained on the \textit{Aliyun}, \textit{GraphMal}, and \textit{Aliyun+Catak} datasets, respectively, and tested on the \textit{GraphMal}, \textit{Aliyun}, and \textit{GraphMal} datasets, respectively.}
		\label{fewshot}
	\end{figure*}

	\subsection{Comparison of Few-shot Learning}
	A straightforward approach to few-shot learning involves fine-tuning a model on a support set, with this model being based on one that has already undergone training. The model is then evaluated through a query set\cite{chai2022dynamic,tran2020mannware,wang2021novel}. However, in practical applications, due to the significant data drift phenomenon within malware samples, the model trained in the present may not yield satisfactory predictive results on future malware. 

In the few-shot learning experiment, different datasets are utilized for training and testing. Additionally, within these testing datasets, a limited number of samples are employed to fine-tune the trained model. As illustrated in Figure~\ref{fewshot}, the proposed model converges more quickly and shows superior fine-tuning in comparison to both TextCNN and BiLSTM.

Though the representation yielded by the proposed method is frozen, it has both high quality and excellent generalization. Consequently, it is only necessary to adjust subsequent module parameters to adapt to the new dataset distribution expediently. When encountering new samples, TextCNN and BiLSTM parameters within the representation layer require adjustment to adapt to the updated dataset distribution. However, because of the limited sample size, the quality of the representation adjustment is not high, subsequently impairing their fine-tuning performance.

	\subsection{Comparison of Explanatory Text Acquisition}
In order to measure the quality of the explanatory text generated by our proposed method, we employ two explanatory text acquisition methods for comparison:	
	The \textbf{D}ocument \textbf{R}etrieval (\textbf{DR}) method seeks API calls and collects their meanings from the Windows API reference manual published by the Office Training Center in China \footnote{http://www.office-cn.net/t/api/index.html?web.htm}. This reference manual includes explanations and parameter interpretations for common API calls.
	The \textbf{I}nternet \textbf{S}earch (\textbf{IS}) method manually searches for API calls on the internet to yield explanations or introductions. The results of these comparisons are presented in Table ~\ref{Explanatory}.
        	
		\begin{table}[htbp]
		\caption{The performance comparison among various explanatory text acquisition methods is discussed in this section. The term \textbf{Missing Rate} refers to the proportion of API calls that are unsuccessful in obtaining explanatory text, relative to the total number of API calls. Meanwhile, \textbf{ACC} signifies the outcomes of the validation performed on the \textit{Aliyun} dataset.}
		\adjustbox{center}{
			\resizebox{0.4\textwidth}{!}{
				\renewcommand{\arraystretch}{1.3}	
				\begin{tabular}{rccccc}
					\Xhline{3\arrayrulewidth}
					\textbf{Method} & \textbf{Missing Rate$\downarrow
						$} & \textbf{Length} & \textbf{ACC$\uparrow
						$} \\ \hline
					IS              & 40.79\%               & 12.86           & 81.79\%                   \\
					DR              & 78.62\%               & 244.6           & 83.15\%                   \\
					Proposed        & 0\%                   & 93.7            & 85.89\% \\ \Xhline{3\arrayrulewidth}
				\end{tabular}
			}
		}
		\label{Explanatory}
	\end{table}
	
For certain API calls, the explanatory text cannot be obtained using the DR and IS methods, resulting in a higher missing rate compared to the proposed method. Additionally, the explanatory text from the DR method is too long while the IS method produces overly short text, leading to moderate detection results. The proposed method, however, can generate corresponding explanatory texts for all API calls. Due to the profound knowledge storage of the \textit{GPT-4}, the explanatory texts are of high quality. The word count is efficiently controlled, leading to improved detection performance. The proposed method also dramatically lowers both manpower and time costs necessary for explanatory text retrieval and eliminates the need for text preprocessing operations.

\begin{figure}[t]
		
		\begin{minipage}{0.48\textwidth}
			\centering
			\includegraphics[width=0.9\linewidth, height=0.6\linewidth]{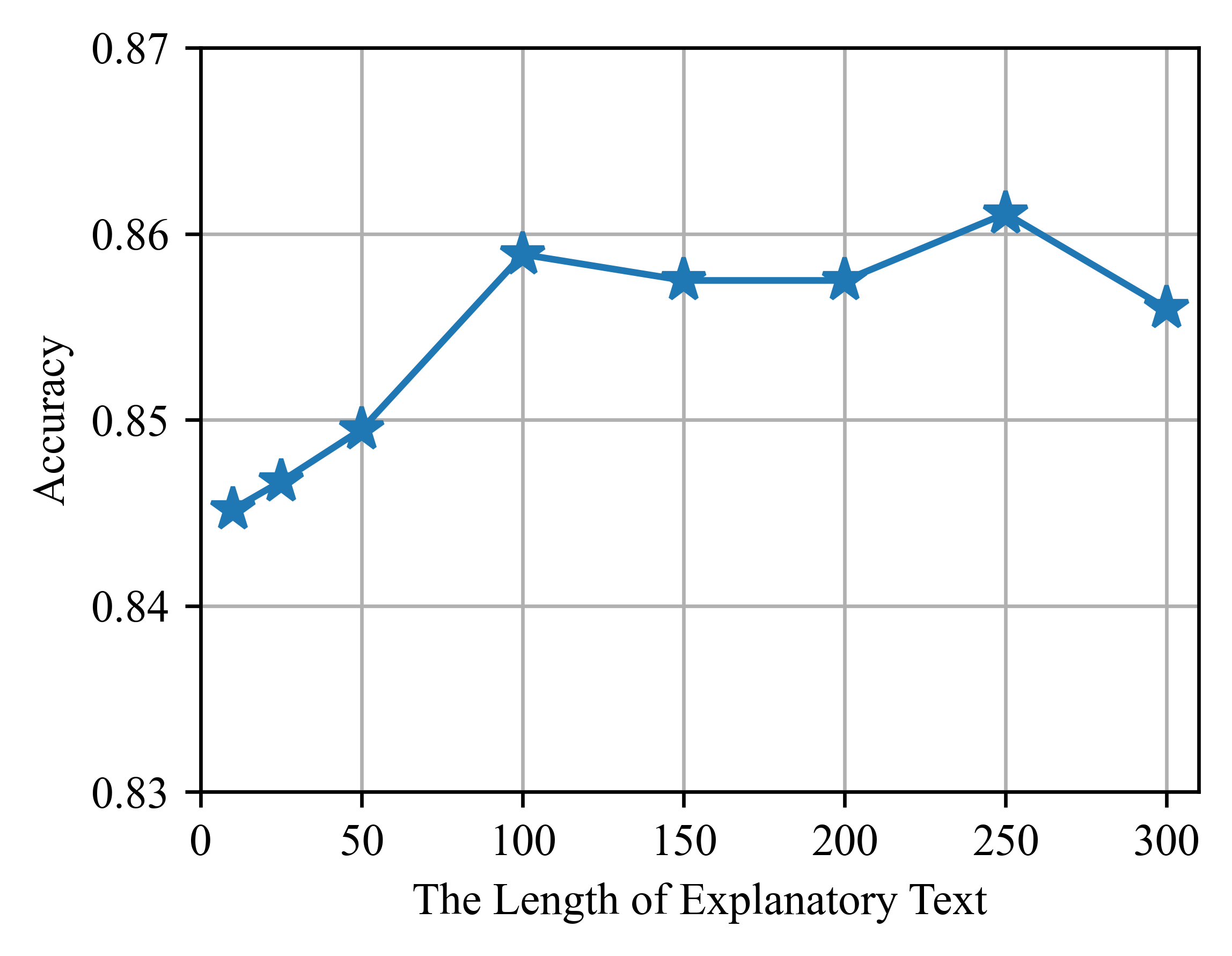} 
			\caption{The impact of explanatory text length on the model's detection performance.}
			\label{Text}
		\end{minipage}
		
	\end{figure}

\begin{figure}[t]
	
	\begin{minipage}{0.5\textwidth}
		\centering
		\includegraphics[width=0.95\linewidth, height=0.58\linewidth]{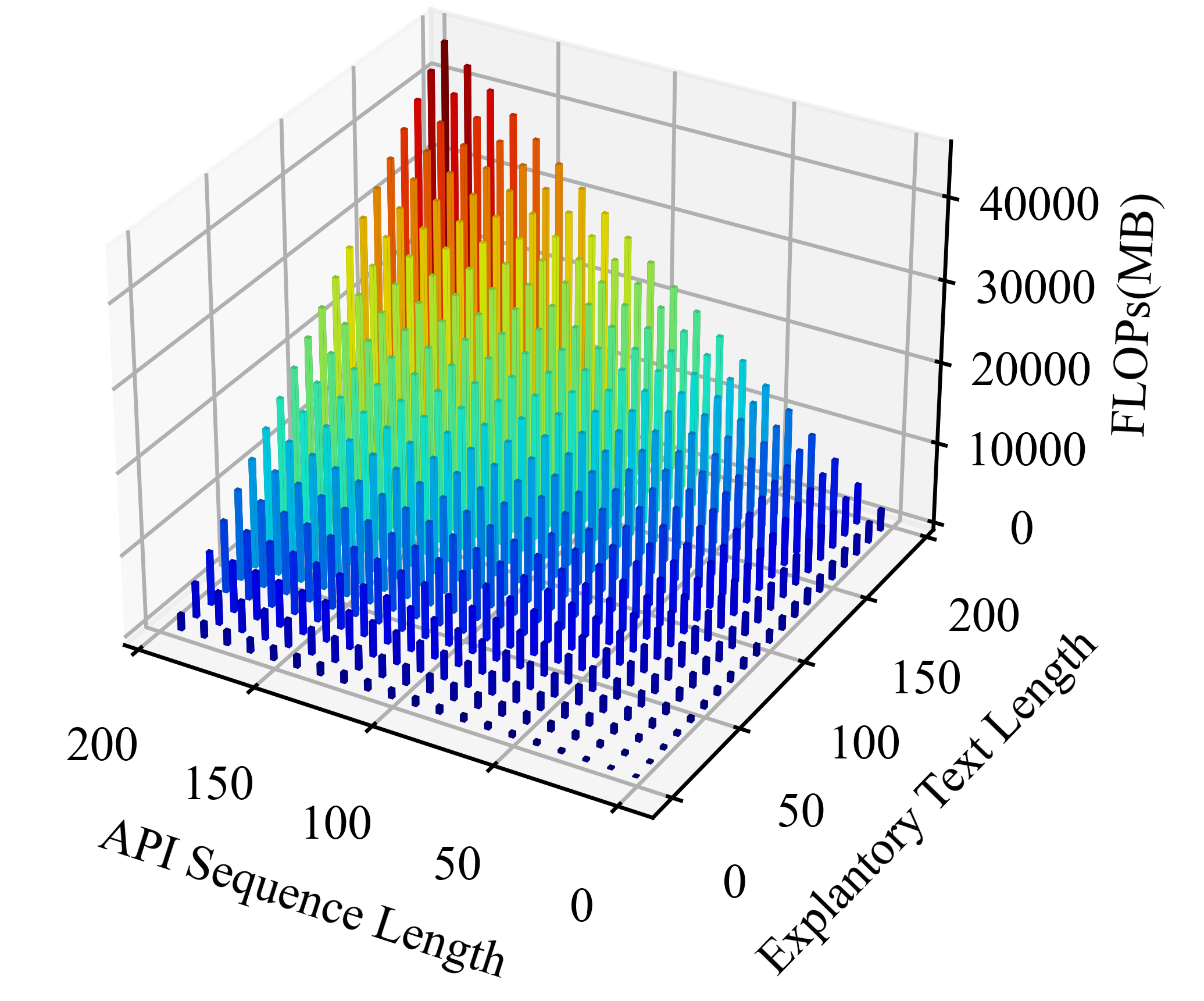} 
		\caption{The correlation among FLOPs, API sequence length and explanatory text length. As the length of the API sequence and explanatory text increases, FLOPs markedly rise, necessitating more GPU memory space for storage, along with an increase in computation time.}
		\label{FLOPs}
	\end{minipage}
	
\end{figure}

The relationship between the length of the explanatory text and the model's performance is explored, as depicted in Figure~\ref{Text}. If the explanatory text is too short, it may not adequately describe the API calls. Conversely, if the explanatory text is too long, it could reach a point of saturation in accuracy where further increases in text length do not improve detection performance. Instead, it uses up more computational time and space (Figure~\ref{FLOPs}). Overly long texts could introduce redundant information potentially deteriorating the model's detection performance. It's worth noting that changes in the length of the explanatory text do not cause significant fluctuations in detection performance. Therefore, the length of the explanatory text is not a sensitive parameter.

	\begin{table*}[htb]
	\centering
	\caption{For identical API calls, explanations of both the earlier version and the current version are provided by \textit{GPT-4}.}
	\adjustbox{center}{
		\resizebox{1.0\textwidth}{!}{
			\renewcommand{\arraystretch}{1.3}	
			\begin{tabular}{c|p{7cm}|p{7cm}}  
				\Xhline{3\arrayrulewidth}
				\textbf{API Call}       & \textbf{Earlier Version                      }                                                                                                                                                                                                                                                                                                                  & \textbf{Current Version  }                                                                                                                                                                                                                                                                                                                                                                                                                                                                                                                                                                  \\ \hline
				{\tt{GetVersionEx}}  & In early versions of Windows, the `GetVersionEx` function could be used to obtain detailed operating system version information, including the major version number, minor version number, build number, platform ID, and additional version information (returned via other members of the `OSVERSIONINFOEX` structure).                              & However, starting with Windows 8.1 and Windows Server 2012 R2, the behavior of the `GetVersionEx` function changed. If an application that calls `GetVersionEx` does not have a manifest declaring its compatibility with Windows 8.1 or higher, then the function will return version information for the highest version of Windows with which the application is compatible, rather than the actual version of the operating system on which it is running. This is because Microsoft wants to encourage developers to program for features, not for operating system versions. \\ \hline
				{\tt{CreateWindowEx}} & In earlier versions, it was primarily used to create a window with specified styles, name, position, and size. However, over time, the functionality of the CreateWindowEx function has been expanded and it now includes more parameters and options to support more complex window creation requirements.                                            & The CreateWindowEx function has added some new parameters, such as an extended window style parameter (dwExStyle). This parameter allows developers to set some advanced window styles, such as transparent windows, tool windows, and windows with shadows. This means that the modern CreateWindowEx function offers more flexibility and developers can use it to create more complex windows.                                                                                                                                                                                  \\ \hline
				{\tt{CreateProcess}}  & In Windows XP and earlier versions, the CreateProcess function directly creates a new process from the specified command-line argument. This function does not check whether the executable to be created contains a manifest. A manifest is an XML file that describes one or more assemblies, including name, version number, public key token, etc. & However, starting from Windows Vista, the behavior of the CreateProcess function has changed. Now, when you call the CreateProcess function, it first checks whether the specified executable file has a manifest. If there is a manifest, CreateProcess uses the information in the manifest to create a new process. This may result in a different behavior of the CreateProcess function in newer Windows versions if the executable file contains a manifest.     \\ \Xhline{3\arrayrulewidth}                                                                                                         
			\end{tabular}
		}
	}
	\label{concept}
\end{table*}

\subsection{Analysis of Concept Drift Alleviation}

The phenomenon of data distribution evolution over time, which impacts the detection performance of models, is known as concept drift \cite{lu2014concept,lu2018learning}. An effective method to address this situation is incremental learning. By learning from new data, detection models can recognize evolving data distributions and enhance their capacity to detect novel samples. This phenomenon is particularly noticeable in the realm of malware API call behavior. Changes, such as the introduction of new API calls and updates to existing API calls, can influence the model's detection performance.

In Section~\ref{Domain}, $\mathcal{D}_{large}$ introduces some new API calls compared to $\mathcal{D}_{base}$. With the assistance of \textit{GPT-4}, our method can generate representations for unknown or newly introduced API calls, and achieve an excellent recall rate of malware, thereby demonstrating the ability to handle concept drift to a certain degree. Besides, it can also monitor the latest interpretations of API calls. Given that the knowledge obtained through \textit{GPT-4} is continually updated, this is advantageous for ongoing learning to manage the concept drift phenomenon. Furthermore, it allows the tracking of API call explanations over specific periods, with some examples demonstrated in Table ~\ref{concept}.

	\section{Conclusion}
	This paper proposes a non-training representation generation method using \textit{GPT-4} prompts.
	We first design the prompt text to guide \textit{GPT-4} to generate the explanatory text of each API call, then perform pre-trained BERT to generate the representation of each explanatory text, and finally, a CNN-based module is constructed to learn the representation, thereby achieving excellent detection performance of the proposed model.
	 The generation of this representation is not reliant on malware datasets training and, theoretically, it can generate the representation for all API calls. Consequently, this method can effectively address the issues of weak generalization and concept drift. 
The detection performance, particularly the generation capacity, of our proposed model has seen improvements when compared to SOTA models. 

In future work, we aim to collect more datasets of malware representation and analyze them. Through this, we strive to provide a solid foundation for the creation of a large-scale model specifically designed for malware representation and detection.
	
	\bibliographystyle{IEEEtran}
	\bibliography{NLP,Transformer,GPTData,Easy,Intro}

% Generated by IEEEtran.bst, version: 1.14 (2015/08/26)
\begin{thebibliography}{10}
\providecommand{\url}[1]{#1}
\csname url@samestyle\endcsname
\providecommand{\newblock}{\relax}
\providecommand{\bibinfo}[2]{#2}
\providecommand{\BIBentrySTDinterwordspacing}{\spaceskip=0pt\relax}
\providecommand{\BIBentryALTinterwordstretchfactor}{4}
\providecommand{\BIBentryALTinterwordspacing}{\spaceskip=\fontdimen2\font plus
\BIBentryALTinterwordstretchfactor\fontdimen3\font minus
  \fontdimen4\font\relax}
\providecommand{\BIBforeignlanguage}[2]{{%
\expandafter\ifx\csname l@#1\endcsname\relax
\typeout{** WARNING: IEEEtran.bst: No hyphenation pattern has been}%
\typeout{** loaded for the language `#1'. Using the pattern for}%
\typeout{** the default language instead.}%
\else
\language=\csname l@#1\endcsname
\fi
#2}}
\providecommand{\BIBdecl}{\relax}
\BIBdecl

\bibitem{guizani2020network}
N.~Guizani and A.~Ghafoor, ``A network function virtualization system for
  detecting malware in large iot based networks,'' \emph{IEEE Journal on
  Selected Areas in Communications}, vol.~38, no.~6, pp. 1218--1228, 2020.

\bibitem{amira2023survey}
A.~Amira, A.~Derhab, E.~B. Karbab, and O.~Nouali, ``A survey of malware
  analysis using community detection algorithms,'' \emph{ACM Computing
  Surveys}, vol.~56, no.~2, pp. 1--29, 2023.

\bibitem{gopinath2023comprehensive}
M.~Gopinath and S.~C. Sethuraman, ``A comprehensive survey on deep learning
  based malware detection techniques,'' \emph{Computer Science Review},
  vol.~47, p. 100529, 2023.

\bibitem{uppal2014malware}
D.~Uppal, R.~Sinha, V.~Mehra, and V.~Jain, ``Malware detection and
  classification based on extraction of api sequences,'' in \emph{2014
  International Conference on Advances in Computing, Communications and
  Informatics (ICACCI)}.\hskip 1em plus 0.5em minus 0.4em\relax IEEE, 2014, pp.
  2337--2342.

\bibitem{pascanu2015malware}
R.~Pascanu, J.~W. Stokes, H.~Sanossian, M.~Marinescu, and A.~Thomas, ``Malware
  classification with recurrent networks,'' in \emph{2015 IEEE International
  Conference on Acoustics, Speech and Signal Processing (ICASSP)}.\hskip 1em
  plus 0.5em minus 0.4em\relax IEEE, 2015, pp. 1916--1920.

\bibitem{athiwaratkun2017malware}
B.~Athiwaratkun and J.~W. Stokes, ``Malware classification with {LSTM} and
  {GRU} language models and a character-level {CNN},'' in \emph{2017 IEEE
  International Conference on Acoustics, Speech and Signal Processing
  (ICASSP)}.\hskip 1em plus 0.5em minus 0.4em\relax IEEE, 2017, pp. 2482--2486.

\bibitem{maniath2017deep}
S.~Maniath, A.~Ashok, P.~Poornachandran, V.~Sujadevi, P.~S. AU, and S.~Jan,
  ``Deep learning {LSTM} based ransomware detection,'' in \emph{2017 Recent
  Developments in Control, Automation \& Power Engineering (RDCAPE)}.\hskip 1em
  plus 0.5em minus 0.4em\relax IEEE, 2017, pp. 442--446.

\bibitem{catak2020deep}
F.~O. Catak, A.~F. Yaz{\i}, O.~Elezaj, and J.~Ahmed, ``Deep learning based
  sequential model for malware analysis using windows exe {API} calls,''
  \emph{PeerJ Computer Science}, vol.~6, p. e285, 2020.

\bibitem{vaswani2017attention}
A.~Vaswani, N.~Shazeer, N.~Parmar, J.~Uszkoreit, L.~Jones, A.~N. Gomez,
  {\L}.~Kaiser, and I.~Polosukhin, ``Attention is all you need,''
  \emph{Advances in neural information processing systems}, vol.~30, 2017.

\bibitem{devlin2018bert}
J.~Devlin, M.-W. Chang, K.~Lee, and K.~Toutanova, ``Bert: Pre-training of deep
  bidirectional transformers for language understanding,'' \emph{arXiv preprint
  arXiv:1810.04805}, 2018.

\bibitem{radford2019language}
A.~Radford, J.~Wu, R.~Child, D.~Luan, D.~Amodei, I.~Sutskever \emph{et~al.},
  ``Language models are unsupervised multitask learners,'' \emph{OpenAI blog},
  vol.~1, no.~8, p.~9, 2019.

\bibitem{radford2021learning}
A.~Radford, J.~W. Kim, C.~Hallacy, A.~Ramesh, G.~Goh, S.~Agarwal, G.~Sastry,
  A.~Askell, P.~Mishkin, J.~Clark \emph{et~al.}, ``Learning transferable visual
  models from natural language supervision,'' in \emph{International conference
  on machine learning}.\hskip 1em plus 0.5em minus 0.4em\relax PMLR, 2021, pp.
  8748--8763.

\bibitem{yao2024PromptCARE}
H.~Yao, J.~Lou, K.~Ren, and Z.~Qin, ``Promptcare: Prompt copyright protection
  by watermark injection and verification,'' in \emph{IEEE Symposium on
  Security and Privacy (S\&P)}.\hskip 1em plus 0.5em minus 0.4em\relax IEEE,
  2024.

\bibitem{demirkiran2022ensemble}
F.~Demirk{\i}ran, A.~{\c{C}}ay{\i}r, U.~{\"U}nal, and H.~Da{\u{g}}, ``An
  ensemble of pre-trained transformer models for imbalanced multiclass malware
  classification,'' \emph{Computers \& Security}, vol. 121, p. 102846, 2022.

\bibitem{xu2021malbert}
Z.~Xu, X.~Fang, and G.~Yang, ``Malbert: A novel pre-training method for malware
  detection,'' \emph{Computers \& Security}, vol. 111, p. 102458, 2021.

\bibitem{GPT4}
OpenAI, ``{GPT-4} technical report,'' \emph{arXiv preprint arXiv:2303.08774},
  2022.

\bibitem{chowdhery2022palm}
A.~Chowdhery, S.~Narang, J.~Devlin, M.~Bosma, G.~Mishra, A.~Roberts, P.~Barham,
  H.~W. Chung, C.~Sutton, S.~Gehrmann \emph{et~al.}, ``Palm: Scaling language
  modeling with pathways,'' \emph{arXiv preprint arXiv:2204.02311}, 2022.

\bibitem{touvron2023llama}
H.~Touvron, T.~Lavril, G.~Izacard, X.~Martinet, M.-A. Lachaux, T.~Lacroix,
  B.~Rozi{\`e}re, N.~Goyal, E.~Hambro, F.~Azhar \emph{et~al.}, ``Llama: Open
  and efficient foundation language models,'' \emph{arXiv preprint
  arXiv:2302.13971}, 2023.

\bibitem{chatglm}
Z.~Du, Y.~Qian, X.~Liu, M.~Ding, J.~Qiu, Z.~Yang, and J.~Tang, ``Glm: General
  language model pretraining with autoregressive blank infilling,'' in
  \emph{the 60th Annual Meeting of the Association for Computational
  Linguistics}, 2022, pp. 320--335.

\bibitem{alazab2010towards}
M.~Alazab, S.~Venkataraman, and P.~Watters, ``Towards understanding malware
  behaviour by the extraction of api calls,'' in \emph{2010 Second Cybercrime
  and Trustworthy Computing Workshop}.\hskip 1em plus 0.5em minus 0.4em\relax
  IEEE, 2010, pp. 52--59.

\bibitem{gupta2016malware}
S.~Gupta, H.~Sharma, and S.~Kaur, ``Malware characterization using windows api
  call sequences,'' in \emph{Security, Privacy, and Applied Cryptography
  Engineering: 6th International Conference}.\hskip 1em plus 0.5em minus
  0.4em\relax Springer, 2016, pp. 271--280.

\bibitem{ravi2012malware}
C.~Ravi and R.~Manoharan, ``Malware detection using windows api sequence and
  machine learning,'' \emph{International Journal of Computer Applications},
  vol.~43, no.~17, pp. 12--16, 2012.

\bibitem{ki2015novel}
Y.~Ki, E.~Kim, and H.~K. Kim, ``A novel approach to detect malware based on api
  call sequence analysis,'' \emph{International Journal of Distributed Sensor
  Networks}, vol.~11, no.~6, p. 659101, 2015.

\bibitem{sami2010malware}
A.~Sami, B.~Yadegari, H.~Rahimi, N.~Peiravian, S.~Hashemi, and A.~Hamze,
  ``Malware detection based on mining api calls,'' in \emph{the 2010 ACM
  Symposium on Applied Computing}, 2010, pp. 1020--1025.

\bibitem{pektacs2018malware}
A.~Pekta{\c{s}} and T.~Acarman, ``Malware classification based on api calls and
  behaviour analysis,'' \emph{IET Information Security}, vol.~12, no.~2, pp.
  107--117, 2018.

\bibitem{anderson2011graph}
B.~Anderson, D.~Quist, J.~Neil, C.~Storlie, and T.~Lane, ``Graph-based malware
  detection using dynamic analysis,'' \emph{Journal in computer Virology},
  vol.~7, pp. 247--258, 2011.

\bibitem{shijo2015integrated}
P.~Shijo and A.~Salim, ``Integrated static and dynamic analysis for malware
  detection,'' \emph{Procedia Computer Science}, vol.~46, pp. 804--811, 2015.

\bibitem{islam2013classification}
R.~Islam, R.~Tian, L.~M. Batten, and S.~Versteeg, ``Classification of malware
  based on integrated static and dynamic features,'' \emph{Journal of Network
  and Computer Applications}, vol.~36, no.~2, pp. 646--656, 2013.

\bibitem{yuan2020character}
L.~Yuan, Z.~Zeng, Y.~Lu, X.~Ou, and T.~Feng, ``A character-level
  {BiGRU}-attention for phishing classification,'' in \emph{Information and
  Communications Security: 21st International Conference, ICICS 2019}.\hskip
  1em plus 0.5em minus 0.4em\relax Springer, 2020, pp. 746--762.

\bibitem{dang2021malware}
D.~Dang, F.~Di~Troia, and M.~Stamp, ``Malware classification using long
  short-term memory models,'' \emph{arXiv preprint arXiv:2103.02746}, 2021.

\bibitem{qin2020api}
B.~Qin, Y.~Wang, and C.~Ma, ``{API} call based ransomware dynamic detection
  approach using textcnn,'' in \emph{2020 International Conference on Big Data,
  Artificial Intelligence and Internet of Things Engineering (ICBAIE)}.\hskip
  1em plus 0.5em minus 0.4em\relax IEEE, 2020, pp. 162--166.

\bibitem{kim2014convolutional}
Y.~Kim, ``Convolutional neural networks for sentence classification,'' in
  \emph{the 2014 Conference on Empirical Methods in Natural Language
  Processing, {EMNLP} 2014}.\hskip 1em plus 0.5em minus 0.4em\relax {ACL},
  2014, pp. 1746--1751.

\bibitem{li2022novel}
C.~Li, Q.~Lv, N.~Li, Y.~Wang, D.~Sun, and Y.~Qiao, ``A novel deep framework for
  dynamic malware detection based on {API} sequence intrinsic features,''
  \emph{Computers \& Security}, vol. 116, p. 102686, 2022.

\bibitem{brown2020language}
T.~B. Brown, B.~Mann, N.~Ryder, M.~Subbiah, J.~Kaplan, P.~Dhariwal,
  A.~Neelakantan, P.~Shyam, G.~Sastry, A.~Askell, S.~Agarwal,
  A.~Herbert{-}Voss, G.~Krueger, T.~Henighan, R.~Child, A.~Ramesh, D.~M.
  Ziegler, J.~Wu, C.~Winter, C.~Hesse, M.~Chen, E.~Sigler, M.~Litwin, S.~Gray,
  B.~Chess, J.~Clark, C.~Berner, S.~McCandlish, A.~Radford, I.~Sutskever, and
  D.~Amodei, ``Language models are few-shot learners,'' in \emph{Advances in
  Neural Information Processing Systems}, 2020.

\bibitem{raffel2020exploring}
C.~Raffel, N.~Shazeer, A.~Roberts, K.~Lee, S.~Narang, M.~Matena, Y.~Zhou,
  W.~Li, and P.~J. Liu, ``Exploring the limits of transfer learning with a
  unified text-to-text transformer,'' \emph{The Journal of Machine Learning
  Research}, vol.~21, no.~1, pp. 5485--5551, 2020.

\bibitem{dai2019transformer}
Z.~Dai, Z.~Yang, Y.~Yang, J.~G. Carbonell, Q.~V. Le, and R.~Salakhutdinov,
  ``Transformer-xl: Attentive language models beyond a fixed-length context,''
  in \emph{the 57th Conference of the Association for Computational
  Linguistics}.\hskip 1em plus 0.5em minus 0.4em\relax Association for
  Computational Linguistics, 2019, pp. 2978--2988.

\bibitem{yang2019xlnet}
Z.~Yang, Z.~Dai, Y.~Yang, J.~G. Carbonell, R.~Salakhutdinov, and Q.~V. Le,
  ``Xlnet: Generalized autoregressive pretraining for language understanding,''
  in \emph{Advances in Neural Information Processing Systems}, H.~M. Wallach,
  H.~Larochelle, A.~Beygelzimer, F.~d'Alch{\'{e}}{-}Buc, E.~B. Fox, and
  R.~Garnett, Eds., 2019, pp. 5754--5764.

\bibitem{rahali2021malbert}
A.~Rahali and M.~A. Akhloufi, ``Malbert: Malware detection using bidirectional
  encoder representations from transformers,'' in \emph{2021 IEEE International
  Conference on Systems, Man, and Cybernetics (SMC)}.\hskip 1em plus 0.5em
  minus 0.4em\relax IEEE, 2021, pp. 3226--3231.

\bibitem{demirci2022static}
D.~Dem{\i}rc{\i}, C.~Acarturk \emph{et~al.}, ``Static malware detection using
  stacked bilstm and gpt-2,'' \emph{IEEE Access}, vol.~10, pp.
  58\,488--58\,502, 2022.

\bibitem{rahali2023malbertv2}
A.~Rahali and M.~A. Akhloufi, ``Malbertv2: Code aware bert-based model for
  malware identification,'' \emph{Big Data and Cognitive Computing}, vol.~7,
  no.~2, p.~60, 2023.

\bibitem{ferrag2023revolutionizing}
M.~A. Ferrag, M.~Ndhlovu, N.~Tihanyi, L.~C. Cordeiro, M.~Debbah, and
  T.~Lestable, ``Revolutionizing cyber threat detection with large language
  models,'' \emph{arXiv preprint arXiv:2306.14263}, 2023.

\bibitem{wei2022chain}
J.~Wei, X.~Wang, D.~Schuurmans, M.~Bosma, F.~Xia, E.~Chi, Q.~V. Le, D.~Zhou
  \emph{et~al.}, ``Chain-of-thought prompting elicits reasoning in large
  language models,'' \emph{Advances in Neural Information Processing Systems},
  vol.~35, pp. 24\,824--24\,837, 2022.

\bibitem{yang2023dawn}
Z.~Yang, L.~Li, K.~Lin, J.~Wang, C.-C. Lin, Z.~Liu, and L.~Wang, ``The dawn of
  lmms: Preliminary explorations with gpt-4v (ision),'' \emph{arXiv preprint
  arXiv:2309.17421}, 2023.

\bibitem{zhang2020dynamic}
Z.~Zhang, P.~Qi, and W.~Wang, ``Dynamic malware analysis with feature
  engineering and feature learning,'' in \emph{The Thirty-Fourth {AAAI}
  Conference on Artificial Intelligence, {AAAI} 2020}, vol.~34, no.~01, 2020,
  pp. 1210--1217.

\bibitem{kolosnjaji2016deep}
B.~Kolosnjaji, A.~Zarras, G.~Webster, and C.~Eckert, ``Deep learning for
  classification of malware system call sequences,'' in \emph{Advances in
  Artificial Intelligence: 29th Australasian Joint Conference}.\hskip 1em plus
  0.5em minus 0.4em\relax Springer International Publishing, 2016, pp.
  137--149.

\bibitem{zhang2023dynamic}
S.~Zhang, J.~Wu, M.~Zhang, and W.~Yang, ``Dynamic malware analysis based on api
  sequence semantic fusion,'' \emph{Applied Sciences}, vol.~13, no.~11, p.
  6526, 2023.

\bibitem{dataset:mal_sandbox}
\BIBentryALTinterwordspacing
{Alibaba Cloud}, ``Alibaba cloud malware detection based on behaviors,'' 2018,
  [Online; accessed 11-November-2018]. [Online]. Available:
  \url{https://tianchi.aliyun.com/getStart/information.htm?raceId=231694}
\BIBentrySTDinterwordspacing

\bibitem{oliveira2019behavioral}
A.~Oliveira and R.~Sassi, ``Behavioral malware detection using deep graph
  convolutional neural networks,'' \emph{TechRxiv}, p. preprint, 2019.

\bibitem{Virus}
\BIBentryALTinterwordspacing
khas ccip, ``Api sequences malware datasets,'' 2021, 2023-10. [Online].
  Available: \url{https://github.com/khas-ccip/api_sequences_malware_datasets}
\BIBentrySTDinterwordspacing

\bibitem{chai2022dynamic}
Y.~Chai, L.~Du, J.~Qiu, L.~Yin, and Z.~Tian, ``Dynamic prototype network based
  on sample adaptation for few-shot malware detection,'' \emph{IEEE
  Transactions on Knowledge and Data Engineering}, vol.~35, no.~5, pp.
  4754--4766, 2022.

\bibitem{tran2020mannware}
K.~Tran, H.~Sato, and M.~Kubo, ``Mannware: A malware classification approach
  with a few samples using a memory augmented neural network,''
  \emph{Information}, vol.~11, no.~1, p.~51, 2020.

\bibitem{wang2021novel}
P.~Wang, Z.~Tang, and J.~Wang, ``A novel few-shot malware classification
  approach for unknown family recognition with multi-prototype modeling,''
  \emph{Computers \& Security}, vol. 106, p. 102273, 2021.

\bibitem{lu2014concept}
N.~Lu, G.~Zhang, and J.~Lu, ``Concept drift detection via competence models,''
  \emph{Artificial Intelligence}, vol. 209, pp. 11--28, 2014.

\bibitem{lu2018learning}
J.~Lu, A.~Liu, F.~Dong, F.~Gu, J.~Gama, and G.~Zhang, ``Learning under concept
  drift: A review,'' \emph{IEEE Transactions on Knowledge and Data
  Engineering}, vol.~31, no.~12, pp. 2346--2363, 2018.

\end{thebibliography}

\end{document}